\documentclass[preprint,3p]{elsarticle}

\usepackage{natbib}

\usepackage{graphicx}
\usepackage{url}
\usepackage{framed}
\usepackage{colortbl}
\usepackage{tabularx}
\usepackage{prettyref}

\newrefformat{sec}{Section~\ref{#1}}
\newrefformat{fig}{Figure~\ref{#1}}
\newrefformat{tab}{Table~\ref{#1}}

\definecolor{mygray}{gray}{0.7}
\definecolor{darkblue}{cmyk}{0.7451, 0.2745, 0.1373,0.0392}
\definecolor{Gray}{gray}{0.85}

\newcommand{\ie}{i.\,e.\ }
\newcommand{\eg}{e.\,g.\ }

\journal{Information and Software Technology}

\begin{document}

\begin{frontmatter}

\title{Operationalised Product Quality Models and Assessment: \\ The Quamoco Approach}

\author[us]{Stefan Wagner}\corref{cor1} 
\ead{stefan.wagner@informatik.uni-stuttgart.de}
\author[cqse]{Andreas Goeb}
\author[cqse]{Lars Heinemann} 
\author[iese]{Michael Kl\"as}
\author[iese]{Constanza Lampasona}
\author[tum]{Klaus Lochmann}
\author[jku]{Alois Mayr}
\author[jku]{Reinhold Pl\"osch}
\author[bmw]{Andreas Seidl}
\author[ite]{Jonathan Streit}
\author[iese]{Adam Trendowicz}

\address[us]{Institute of Software Technology, University of Stuttgart, Stuttgart, Germany}
\address[cqse]{CQSE GmbH, Garching, Germany}
\address[iese]{Fraunhofer Institute for Experimental Software Engineering IESE, Kaiserslautern, Germany}
\address[tum]{Institut f\"ur Informatik, Technische Universit\"at M\"unchen, Garching, Germany}
\address[jku]{Department of Business Informatics, Johannes Kepler University Linz, Linz, Austria}
\address[bmw]{BMW AG, Munich, Germany}
\address[ite]{itestra GmbH, Munich, Germany}

\cortext[cor1]{Corresponding author}

\begin{abstract}

\textbf{Context}: Software quality models provide either abstract quality 
characteristics or concrete quality measurements; there is no seamless 
integration of these two aspects. Quality assessment approaches are, hence, also very
specific or remain abstract.
Reasons for this include the complexity of quality and the various quality profiles
in different domains which make it difficult to build operationalised
quality models.

\textbf{Objective:} In the project Quamoco, we developed a comprehensive approach aimed at 
closing this gap. 

\textbf{Method:} 
The project combined constructive research, which involved a broad range of quality experts from academia and industry in workshops, 
sprint work and reviews, with empirical studies. All deliverables within the project were peer-reviewed by two project members from a different area.
Most deliverables were developed in two or three iterations and underwent an evaluation.

\textbf{Results:} 
We contribute a comprehensive quality modelling and assessment approach:
(1) A meta quality model defines the structure of operationalised
quality models. It includes the concept of a product factor, which bridges the gap 
between concrete measurements and abstract quality aspects, and allows modularisation to create
modules for specific domains. 
(2) A largely technology-independent base quality model reduces the effort and 
complexity of building quality models for specific domains. 
For Java and C\# systems, we refined it with about~300 concrete product factors and~500 measures. 
(3) A concrete and comprehensive quality assessment approach makes use of the concepts in the meta-model. 
(4) An empirical evaluation of the above results using real-world software systems showed: 
(a) The assessment results using the base model 
largely match the expectations of experts for the corresponding systems. (b) 
The approach and models are well understood by practitioners and considered to be both consistent and well suited 
for getting an overall view on the quality of a software product. The validity of
the base quality model could not be shown conclusively, however.
(5) The extensive, open-source tool support is in a mature state.
(6) The model for embedded software systems is a proof-of-concept for domain-specific quality models.

\textbf{Conclusion:} We provide a broad basis for the development and application of quality models 
in industrial practice as well as a basis for further extension, validation and comparison with other approaches in research.
\end{abstract}

\begin{keyword}
Quality Model \sep Product Quality \sep Quality Assessment
\end{keyword}

\end{frontmatter}

\section{Introduction}

Despite great efforts in both research and practice, software quality
continues to be controversial and insufficiently understood,
and the quality of software products is often unsatisfactory.
The economic impact is enormous and includes
not only spectacular failures of software~\cite{Leveson.2004} but also increased
maintenance costs~\cite{Jones.2011}, high resource consumption, long test cycles, and waiting times for users.

\subsection{Quality Models -- Benefits and Shortcomings}
Software quality models tackle the issues described above by providing a systematic approach for
modelling quality requirements, analysing and monitoring quality,
and directing quality improvement measures \cite{2009_deissenboeckf_purposes_scenarios}. 
They thus allow ensuring quality early in the development process. 

In practice, however, a gap remains between the abstract quality characteristics described in
quality models such as the current standard ISO/IEC~25010~\cite{2011_iso_standard_25010} 
and concrete measurements and assessments~\cite{2009_wagners_quality_models_practice,AlKilidar.2005}. The quality models describe and structure general 
concepts that can be used to characterise software quality. Most of them, however, cannot be used 
for real-world quality assessment or improvement~\cite{AlKilidar.2005}.
Concrete measurements, on the contrary, often lack the
connection to higher-level quality goals. Thus, they make it difficult
to explain the importance of quality problems to developers or sponsors and
to quantify the potential of quality improvements.

A similar gap also exists for quality assessment methods. 
Effective quality management requires not only a definition of quality 
but also a method for assessing the overall quality of a software product 
based on measured properties. Existing quality models either lack  
assessment support completely or provide procedures that are 
too abstract to be operational (\eg ISO/IEC~25040~\cite{iso25040}).
As a consequence, quality assessment is inhibited and likely to produce 
inconsistent and misleading results.

Finally, the contents of many existing quality models (including quality metrics
or evaluation formulas) are invariable. Software quality, however, is not the same
in all contexts. It depends on the domain, on the technology used and on project constraints.
Most quality engineers want to adapt and customise their quality models~\cite{2009_wagners_quality_models_practice}.

\subsection{Research Objective}

Our aim was to develop and validate operationalised quality models for software 
together with a quality assessment method and tool support to
provide the missing connections between generic descriptions of software quality 
characteristics and specific software analysis and measurement approaches.
As a single operationalised quality model that fits all peculiarities of every software domain would
be extremely large and expensive to develop, we also set the goal of enabling modularised
quality models with a widely applicable base model and various specific extensions.
This also constrained the types of analyses and measurements to include: We included
static analyses and manual reviews, because they are least dependent on the system context. In contrast, dynamic
testing of a system would require specific test cases and execution environments. Furthermore, 
we focused on \emph{product} quality and, hence, product aspects influencing quality,  instead
of on process or people aspects. While we consider the latter aspects important as well, we expected
the product aspects to be easier and more direct to measure.

To achieve these goals, software quality experts from both academia 
and industry in Germany joined forces within the Quamoco
research project. The project consortium consisted of Technische 
Universit\"at M\"unchen, SAP AG, Siemens, Capgemini, Fraunhofer 
IESE, and itestra. 
In total, these partners spent 558 person-months
on the project.

\subsection{Contribution}

This article is an extension of an earlier one~\cite{wagner:icse12} and
 provides six major contributions overall:
First, we developed a meta-model for software quality models. It provides
the structure for the full spectrum from organising quality-related concepts to defining
operational means for assessing their fulfilment in a specific
environment including their modularisation. 
Second, the base quality model instantiates the meta-model and captures
knowledge on how to conduct a basic quality assessment
for different kinds of software. It serves as the basis for more specific quality models. 
We elaborated the base model in depth for the languages Java and C\#.

 Third, we contribute
a clearly defined quality assessment method that is integrated with the
meta-model. Fourth, we performed several validations of the base model
with real software systems that showed the understandability of the model
and the correspondence
of the assessment results with expert opinions. Fifth, we developed
extensive tool support for building and adapting operationalised quality 
models as well as for performing quality assessments.
Sixth and last, we contribute a domain-specific quality model as
an extension to the base model, which shows the usefulness of the modularisation
and separation between base model and specific extensions.
Preliminary versions of the meta-model, the base model, the quality assessment and
the empirical analysis were partly described in the earlier conference paper~\cite{wagner:icse12}.

\subsection{Terms and Definitions}

Here, we introduce the terms used most frequently in this article.
\emph{Quality characteristic} refers to the concept as used in ISO/IEC~9126 and ISO/IEC~25010
to decompose software quality, \eg into \emph{maintainability} or \emph{reliability}. Quality characteristics
can hence be thought of as ``-ilities''. A \emph{quality aspect} is more generally an area of interest 
that contributes to the quality of a software product for a certain stakeholder. In particular, we consider
ISO/IEC's quality characteristics as quality aspects but also allow for other decompositions of software 
quality to get increased precision and expressiveness within a quality model. We use one such alternative 
decomposition in the Quamoco base quality model. Quality aspects are typically not measurable 
or observable directly. While they describe quality on a rather abstract level, 
\emph{product factors} are observable properties of certain entities within a software
product. Product factors as well as their relationships to quality aspects are discussed in detail 
in \prettyref{sec:general-concepts}. Finally, a \emph{quality requirement} is a stakeholder's stated 
desire to have a certain quality aspect or product factor fulfilled by a product.

\subsection{Outline}

The remainder of this article is structured as follows: 
\prettyref{sec:relatedwork} presents an overview of both related work 
and our own previous work in the research area of quality modelling and
assessment. In \prettyref{sec:concepts_meta}, we discuss usage scenarios for quality models, introduce general concepts, and present our meta quality model.
We instantiate this meta-model in \prettyref{sec:base_model} and present
our base quality model for software products written in Java or C\#. 
We describe our approach for assessing software quality with this model in \prettyref{sec:quality_assessment}.
\prettyref{sec:tool_support} describes the tool support for the application
of the approach in practice.
In \prettyref{sec:validation}, we describe the empirical evaluation 
we performed to validate the proposed model and tools.
\prettyref{sec:specific-eqm} presents one domain-specific quality model, which
extends the base model for the specifics of embedded systems. Finally, \prettyref{sec:conclusion} concludes the article and outlines future work.

\section{Related Work}
\label{sec:relatedwork}

Quality models have been a research topic for several decades and a large number of 
quality models have been proposed~\cite{2009_klaes_QM_landscape}. We describe the
predominant hierarchical models, important proposals for richer models, quality
measurement approaches and tools, and a summary of our own preliminary work.

\subsection{Hierarchical Models}

The first published quality models for software date back to the late 1970s, when 
Boehm~et~al.~\cite{1978_boehmb_software_quality} as well as 
McCall, Richards and Walter~\cite{McCall.1977} described quality characteristics and their 
decomposition. The two approaches are similar and use a hierarchical decomposition of the
concept \emph{quality} into quality characteristics such as \emph{maintainability} or \emph{reliability}.
Several variations of these models have appeared over time. One of the more popular ones is the 
FURPS model~\cite{grady87}, which decomposes quality into functionality, usability, reliability,
performance, and supportability. 

This kind of quality model became the basis for the international standard ISO/IEC~9126~\cite{iso9126-1:2001} in 1991. 
It defines a standard decomposition into quality characteristics and suggests a small number of metrics for measuring
them. These metrics do not cover all aspects of quality, however. Hence, the standard does by no means 
completely operationalise quality. The successor to ISO/IEC~9126, ISO/IEC~25010~\cite{2011_iso_standard_25010},
changes a few classifications but keeps the general hierarchical decomposition.

In several proposals, researchers have used metrics to directly measure quality characteristics from or 
similar to ISO/IEC~9126. Franch and Carvallo~\cite{franch03} adapt the ISO/IEC quality model and
assign metrics to measure them for selecting software packages. They stress that they need
to be able to explicitly describe ``relationships between quality entities''. 
Van Zeist and Hendriks~\cite{vanzeist96} also extend the ISO/IEC model and attach measures
such as \emph{average learning time}. Samoladas et al.~\cite{samoladas08} use several of
the quality characteristics of ISO/IEC~9126 and extend and adapt them to open source software. They use
the quality characteristics to aggregate measurements to an ordinal scale.
All these approaches reveal that it is necessary to extend and adapt the ISO/IEC standard.
They also show the difficulty in measuring abstract quality characteristics directly.

Various critics (\eg \cite{deissenb:icsm07,AlKilidar.2005}) point out that the decomposition 
principles used for quality characteristics are often ambiguous. Furthermore, the resulting quality 
characteristics are mostly not specific enough to be  measurable directly. Although the recently 
published successor ISO/IEC~25010 has several improvements, including a measurement reference
model in ISO/IEC~25020, the overall criticism is still valid because detailed measures are still missing. 
Moreover, a survey done by us~\cite{2009_wagners_quality_models_practice,Wagner2010} shows that 
fewer than 28\% of the companies use these standard models and 71\% of them have 
developed their own variants. Hence, there is a need for customisation.

\subsection{Richer Models}

Starting in the 1990s, researchers have been proposing more elaborate ways of decomposing quality 
characteristics and thereby have built richer quality models. Dromey~\cite{Dromey.1995} distinguishes between 
\emph{product components}, which exhibit \emph{quality carrying properties}, and externally 
visible \emph{quality attributes}.

Bansiya and Davis~\cite{Bansiya2002} build on Dromey's model and propose QMOOD,
a quality model for object-oriented designs. They describe several metrics for the design of
components to measure what they call \emph{design properties}. These properties have
an influence on quality attributes. They also mention tool support and describe validations
similar to ours (Section~\ref{sec:validation}).

Bakota et al.~\cite{bakota11} emphasise the probabilistic nature of their quality model and
quality assessments. They introduce \emph{virtual quality attributes}, which are similar to
\emph{product factors} in our quality model (cf.~Section~\ref{sec:concepts_meta}). The quality
model uses only nine low-level measures, which are evaluated and aggregated to probability
distributions. Our experience has been that practitioners have difficulties interpreting such
distributions.

Kitchenham~et~al.~\cite{Kitchenham.1997} acknowledge the need for an explicit 
meta-model to describe the increasingly complex structures of quality models. They also
propose a ``build your own'' method for quality models with their Squid approach.

All richer models show that the complex concept of \emph{quality} needs more structure
in quality models than abstract quality characteristics and metrics. They have, however, 
not established a corresponding quality assessment method, tool support and a general base quality 
model, which are necessary for comprehensively measuring and assessing quality.

\subsection{Quality Measurement and Tool Support}

Although not embedded in an operationalised quality model, a large number of
tools for quality analysis are available. Examples include bug pattern identification (\eg 
FindBugs\footnote{\url{http://findbugs.sf.net}}, PMD\footnote{\url{http://pmd.sf.net‎}},
Gendarme\footnote{\url{http://www.mono-project.com/Gendarme}}, or 
PC-Lint\footnote{\url{http://www.gimpel.com/html/pcl.htm}}), coding convention checkers 
(\eg Checkstyle\footnote{\url{http://checkstyle.sf.net}}), clone detection, 
and architecture analysis. These tools focus on specific 
aspects of software quality and fail to provide comprehensive 
quality assessments. Moreover, there are no explicit and systematic links to a quality model.

Dashboards can use the measurement data of these tools as input  
(\eg SonarQube\footnote{\url{http://www.sonarsqube.org}} or
XRadar\footnote{\url{http://xradar.sf.net}}) to provide a visual overview of the 
quality measurements taken from a software system. Nevertheless, they lack an explicit and well-founded
connection between the metrics used and particular quality aspects. Hence, 
explanations of the impacts of defects on software quality and rationales for 
the used metrics are missing.

Experimental research tools (\eg \cite{Marinescu.2005,Schackmann.2009})
take first steps towards integrating a quality model and an assessment toolkit. 
A more comprehensive approach is taken by the research 
project Squale~\cite{Mordal.2012,MordalManet.2009} which was carried out at the same time~\cite{Bergel.2009} as our project. 
Squale provides an explicit quality model based on~\cite{Laval.2008} for describing 
a hierarchical decomposition of the ISO/IEC~9126 quality characteristics.
Hence, it follows and extends the FCM~model~\cite{McCall.1977} 
by introducing an intermediate level of so-called practices which link measures with criteria and hence with quality factors. 
The model contains formulas for evaluating the associated metric values as well as weighting
the sets of practices and criteria for the respective higher-ranked criteria and factors.
For this, Squale provides tool support for evaluating software products and visualising the results.
A shortcoming of Squale is that its quality model does not explicitly consider Dromey's product components to which
the quality issues could be related. These technical details regarding quality issues at the level of product components
make the quality model more useful in practice as this allows for focused drilldowns of quality issues.
Squale also lacks support for modularisation as well as for managing multiple quality models or
multiple hierarchies of quality characteristics.

The CAST Application Intelligence Platform\footnote{\url{http://www.castsoftware.com}} is
a proprietary system that focuses on automatically determining technical debt based on the
risk of quality deficiencies of business applications. Thus, it measures quality on a technical level
and evaluates the risks of vulnerabilities. To do so, it is based on a simple model that 
pools measures into severity classes and does not allow comprehensive quality analysis. 
Users of CAST are provided with a dashboard to monitor
quality deficiencies and resulting risks. 

However, these tools and approaches do, to a certain extent, follow the same idea
of an integrated quality model and its assessment as we do in this article. Regarding the 
quality models, the main difference 
is that we use a product model for structuring the (technical) product 
factors in the Quamoco quality models. Furthermore, Quamoco 
offers modularisation and an editor for creating, adjusting, and managing quality models for various application domains
in contrast to fixed, predefined models. The Quamoco tool chain allows for 
flexible configuration and integration of measurement tools and even manually collected 
data without any changes in the source code of the tool chain.

\subsection{Preliminary Work}

In prior work, we investigated different ways of describing quality and
classifying metrics, \eg activity-based quality models~\cite{deissenb:icsm07} and 
technical issue classifications~\cite{ploesch_et_al_2009}. Based on the findings, we developed 
a meta-model for quality models and evaluated its expressiveness~\cite{Klas.2010}.
We experimented with different approaches for quality assessments and tool 
support~\cite{Plosch.2008,wagner:ist10,Trend.2010,klaes:emse10,Lochmann.2011}. Based on 
the experience gained, we developed a comprehensive tool chain for quality modelling
and assessment~\cite{deissenboeck2011quamoco}.

We published a shorter, preliminary version of this article~\cite{wagner:icse12} which 
focuses on the base model and on a limited set of validations. We added more detailed
descriptions, explanations and illustrations in various parts of the paper to make the work easier to understand. In particular, we added a detailed description of the approach we
used for developing the base quality model. In addition, we added new contributions with the description of domain-specific
quality models, model adaptation, tool support and additional empirical validations.

\section{Quality Model Concepts}
\label{sec:concepts_meta}

The first challenge in addressing the gap between abstract quality
characteristics and concrete assessments is to formalise the structure for 
operationalised quality models in a suitable meta quality model. 
After describing how we use quality models, we will explain each of the concepts briefly and describe 
which problems they solve. Finally, we will combine the concepts into a meta-model to show the complete picture. These concepts and
the meta-model were developed in three iterations spanning three years with corresponding
evaluations~\cite{Klas.2010} (see also Section~\ref{sec:modeldevelopment}).

\subsection{Usage of Quality Models}

Most commonly, we find quality models reduced to mere reference taxonomies or implicitly
implemented in tools. As explicit and living artefacts, however,
they can capture general knowledge about software quality, accumulate knowledge from
their application in projects, and allow quality engineers to define a common understanding of quality in a specific 
context~\cite{2004_marinescur_oo_analysis,deissenb:icsm07,heitlager07,luckey10}.

We aim to use this knowledge as the basis for quality control. In the quality control loop~\cite{deissenb:softw08}, the quality model is the central element for identifying quality requirements, planning quality assurance activities, assessing the fulfilment of quality requirements, and reworking the software product based on the assessment results. The quality model is useful for defining what quality aspects are relevant, how we can measure them and how we can interpret the measurement data to understand the actual quality of a specific software product. This single source of quality information helps to avoid redundancies and inconsistencies which are typical for quality specifications and guidelines.

The quality model defines preferences regarding quality which we need to tailor to the product to be developed in that it specifies relevant quality aspects (not relevant ones are not included in the model) and the level of their relevance (reflected by the importance weights associated with the quality aspects).

In this article, we will focus on the usage of quality models for quality assessment purposes. In this scenario, a software product is measured according to the quality metrics defined in the quality model, and the measurement data are interpreted according to the utility functions defined in the model and then aggregated along the model's structure to derive utilities for individual quality aspects and for product's overall quality. Finally, the aggregated utilities are mapped onto more intuitive quality assessments, for instance ordinal-scale school grades. Please refer to Section \ref{sec:qualityassessmentmethod} for a detailed description of the quality assessment method and an example of its usage.

Quality assessments may be used in a summative or in a formative way. Using the assessment results in a summative way allows quality engineers to assess a software product's quality at specific quality gates and to initiate appropriate quality improvements. The assessment results of a product can be interpreted on its own but also serve as input for a trend analysis, or they can be compared with those of other products in a benchmarking setting. Furthermore, the results can be used in a formative way to spot software defects in the source code. Additionally, applying continuous quality assessments using a quality model may improve the quality awareness of practitioners as they learn from the quality knowledge contained in the quality model.

\subsection{General Concepts}
\label{sec:general-concepts}

The previous work of all Quamoco partners on quality models, our joint discussions, and 
experiences with earlier versions of the meta-model led us to the basic concept
of a \emph{factor}. A factor expresses a \emph{property} of an \emph{entity}, which is similar to what 
Dromey~\cite{Dromey.1995} calls \emph{quality carrying properties} of \emph{product components}. 
We use \emph{entities} to describe the things that are important for quality and
\emph{properties} for the attributes of the things we are interested in. Because the concept of a
factor is general, we can use it on different levels of abstraction. As we focus on product quality 
and aspects of the product that are relevant for it, the entity will always be a part of the software product. For example,
we have concrete factors such as \emph{cohesion of classes} as well as abstract factors such as \emph{portability of the product}. 
The factor concept as such would be generic enough to cover process or people aspects as well, but these
are beyond the scope of the work in this article.

\begin{figure}[htb]
\centering\includegraphics[width=.6\columnwidth]{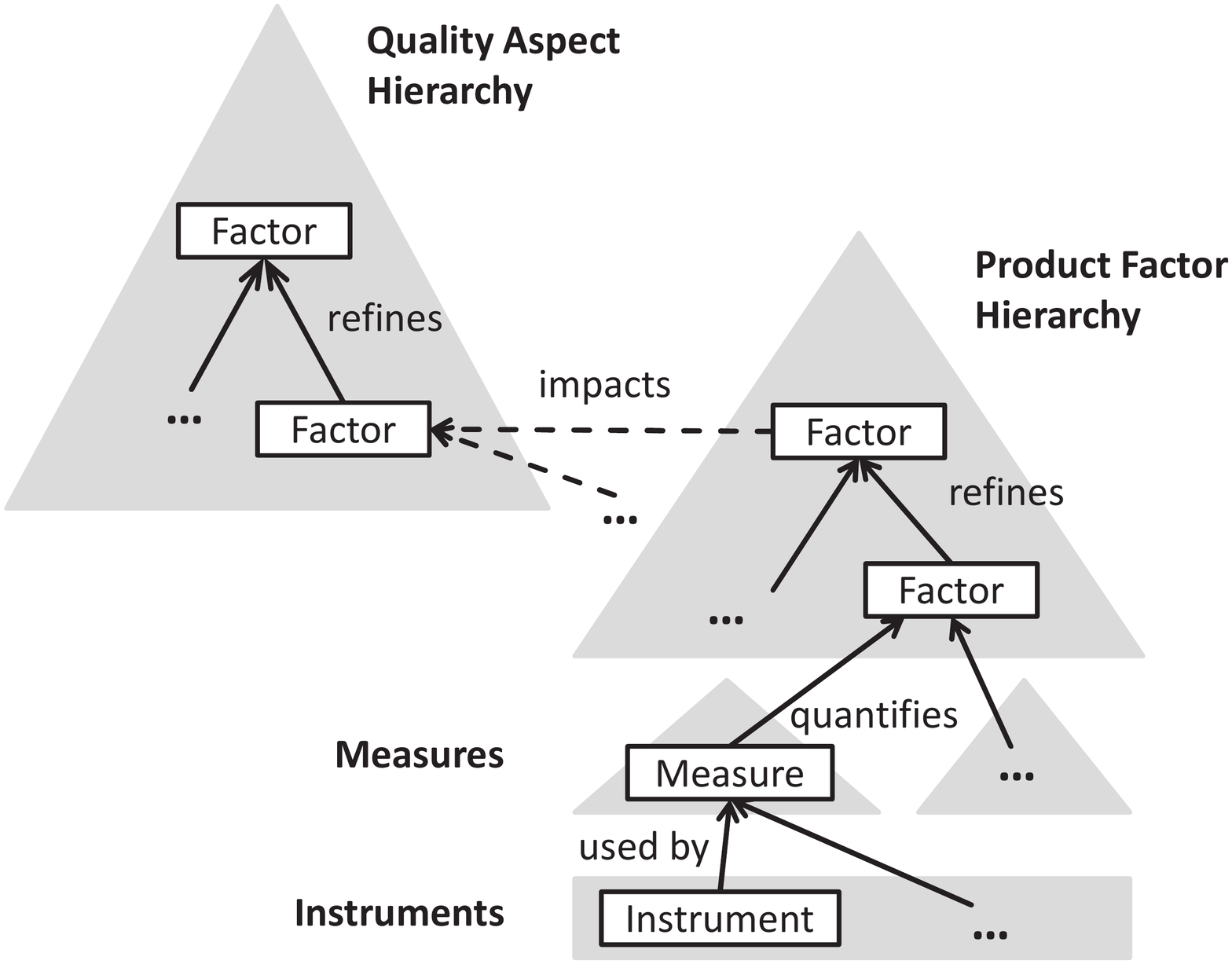}
\caption{The Quality Model Concepts}
\label{fig:model-structure}
\end{figure}

To clearly describe product quality from an abstract level down to concrete measurements, we 
differentiate between the two factor types \emph{quality aspects} and \emph{product factors} which represent
two levels of abstraction. Both can be refined into sub-aspects and sub-factors, respectively, as illustrated in Figure~\ref{fig:model-structure}. 
Quality aspects express abstract quality goals, for example the quality characteristics (or quality attributes) of
ISO/IEC~25010 which always have the complete product as their entity because they describe the whole product.
For example, valid quality aspects are \emph{maintainability of the product}, \emph{analysability of the product} or
\emph{reliability of the product}.

Product factors, in contrast, are attributes of parts of the product.
We require the leaf product factors to be concrete enough to be measured. Examples
are \emph{duplication of source code part}, which we measure with \emph{clone coverage}\footnote{Clone
coverage is the probability that a randomly chosen line of code is duplicated.}, or \emph{detail complexity of method}
measured by length and nesting depth.
This separation of quality aspects and product factors helps us to bridge the gap between abstract notions of quality and concrete
implementations.
In addition, separating the entities from their properties addresses the problem of the difficulty
of decomposing quality attributes. Product factors can easily be decomposed regarding either their property 
or their entity. For example, the entity \emph{class}, representing a part of an object-oriented program, can
be decomposed straightforwardly into the entities \emph{attribute} and \emph{method}. Hence, we can exactly model
for what a property holds.

Moreover, the concepts allow us to model several different hie\-rarchies of quality aspects to
express divergent views on quality. Quality has so many different facets that a single 
hierarchy is not able to express all of them. Even in the recent ISO/IEC~25010, there are two quality
hierarchies: product quality and quality in use. We can model both as quality aspect hierarchies.
Other types of quality aspects are also possible. 
We found that this gives us the flexibility to build quality models tailored, for example for different stakeholders. 
In principle, the concept also allows us 
to have different product factor hierarchies or more levels of abstraction. In our
experiences building quality models, however, we found these two levels to be often sufficient,
given the expressiveness gained from clearly separating entities and properties.

To close the gap between abstract quality aspects and measurable properties of a product, we need to 
put the two factor types into relation. Product factors can have \emph{impacts} 
on quality aspects. This is similar to variation factors, which have impacts on quality factors in
GQM abstraction sheets~\cite{Solingen1999}. An impact is either positive or negative and describes how
the degree of presence or absence of a product factor impacts a quality aspect. This gives us a
complete chain from measured product factors to impacted quality aspects and vice versa. Not all
product factors in our models have impacts because we sometimes use them only for structuring
purposes. Yet, all product factors that we want to have included in assessments need to have
at least one impact on a quality aspect.

Those product factors need to be concrete enough to be measured. Hence, we also have the 
concept of \emph{measures} for product factors.
A measure is a concrete description of how a specific product factor should be quantified
for a specific context. For example, this could be the number of deviations from the rule for Java 
that strings should not be compared using the ``=='' operator or clone coverage as mentioned earlier. 
A product factor can have more than 
one measure if we need multiple measures to cover its concept. 

Moreover, we separate the measures
from their \emph{instruments}. Instruments describe a concrete implementation of a measure. In the
example of the string comparison, an instrument is the corresponding rule as implemented in the
static analysis tool \emph{FindBugs}. This gives us additional flexibility to collect data for measures 
either manually or with different tools in different contexts.

Having these relationships with measures and instruments, it is straightforward to 
assign \emph{evaluations} to product factors and quality aspects to form a quality assessment. The
evaluations aggregate the measurement results (provided by the instruments) for product factors and
the evaluation results of impacting product factors for quality aspects. The intuition is that the evaluations
hold some kind of formula to calculate these aggregations. 
We will describe a comprehensive quality assessment method showing concrete evaluations
in Section~\ref{sec:quality_assessment}. 

Moreover, we
can go the other way round. We can pick quality aspects, for example, ISO/IEC~25010 quality characteristics,
which we consider important and costly for a specific software system, and trace what product factors
affect them and what measures quantify them (cf.~\cite{wagner:ist10}). This allows us to
put emphasis on the product factors with the largest impact on these quality aspects. It also gives us
the basis for specifying quality requirements for which we developed an explicit
quality requirements method~\cite{Ploesch_et_al_2010,Lochmann2010}.

Building quality models with all these element types results in large models with hundreds of model elements.
Not all elements are important in each context and it is impractical to build a single quality model
that contains all measures for all relevant technologies. Therefore, we introduced a modularisation
concept which allows us to split the quality model into \emph{modules}. For example, in the concrete
models described in this article, there is the \emph{root} 
module containing general quality aspect hierarchies as well as basic product factors and
measures. We add additional modules for specific technologies and paradigms, such as
object orientation, programming languages, such as C\#, and domains, such as embedded systems.
This gives us a flexible way to build large and concrete quality models that fit together, meaning they 
are based on the same properties and entities. 

Modularisation enables us to choose appropriate modules and extend the quality model
with additional modules for a given context. To adapt the quality model for a specific company or
project, however, this is still too coarse-grained. Hence, we also developed an explicit adaptation
method, which guides a quality manager in choosing relevant quality aspects, product
factors and measures for a particular project (see Section~\ref{sec:adaptation}).

\subsection{Meta-Model}
\label{sec:meta-model}

\begin{figure}[htb]
\centering\includegraphics[width=.7\columnwidth]{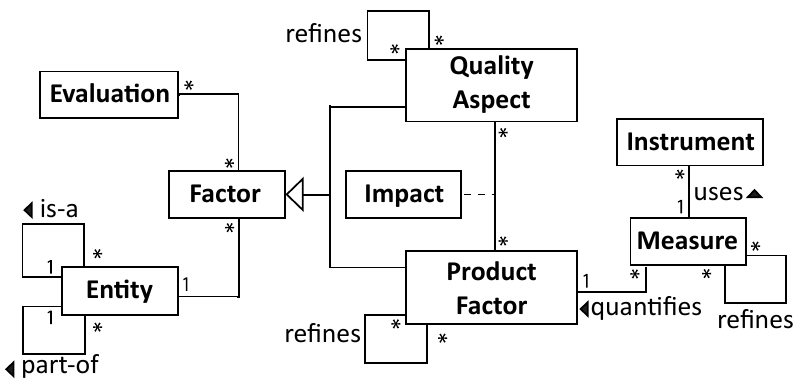}
\caption{The Meta Quality Model (Excerpt)}
\label{fig:metamodel}
\end{figure}

We precisely specified the general concepts described so far in a
meta-model. The core elements of the meta-model are depicted as an (abstracted) UML 
class diagram in Figure~\ref{fig:metamodel}.
Please note that we left out a lot of details such as the IDs, names and descriptions of each
element to make it more comprehensible. A detailed description of the meta-model
is available as a technical report~\cite{Wagner.2012x}. The central element of the meta-model is the \emph{Factor} with
its specialisations \emph{Quality Aspect} and \emph{Product Factor}. Both can be refined and, hence, produce
separate directed acyclic graphs. An \emph{Impact} can only exist between a \emph{Product Factor} and a \emph{Quality Aspect}.
This represents our main relationship between factors and allows us to specify the core quality concepts.

A \emph{Factor} always has an associated \emph{Entity}, which can be in an is-a as well as a
part-of hierarchy. The property of an \emph{Entity} that the \emph{Factor} describes is expressed
in the \emph{Factor}'s name. Each \emph{Factor} may also have an associated \emph{Evaluation}. It specifies how
to evaluate the \emph{Factor}. For that, we can use the evaluation results from sub-factors
or -- in the case of a \emph{Product Factor} -- the values of associated \emph{Measure}s. To keep the UML class diagram
readable, we omit these relationships there.
A \emph{Measure} can be associated with more than one \emph{Product Factor} and has potentially several
\emph{Instruments} that allow us to collect a value for the \emph{Measure} in different contexts, \eg with a
manual inspection or a static analysis tool.

We modelled this meta-model with all details as an EMF\footnote{Eclipse Modeling Framework,
\url{http://emf.eclipse.org/}} model which then served as the basis for the quality model editor
(see Section~\ref{sec:tool_support}).

\section{Base Model} 
\label{sec:base_model}

The base quality model's main objective is to be an operationalised quality
model that can be used directly to assess a wide range of software products
and can also be extended for specific contexts with little effort. To reach this 
goal, the Quamoco project partners conducted several workshops and sprints 
to collaboratively transfer their knowledge and experience into 
the structure described in \prettyref{sec:concepts_meta}.

The resulting quality model represents our consolidated view on
the basics of the quality of software source code and is, in principle, applicable 
to any kind of software. It details quality down to measures and instruments for 
the assessment of Java and C\# systems and, 
hence, enables comprehensive, tool-supported quality assessment without 
requiring large adaptation or configuration effort.

\subsection{Development of the Base Model}
\label{sec:modeldevelopment}

The development of the base model was a joint effort of all
project partners in Quamoco. Overall, at least 23 people from industry and
academia were involved in building the base model. We worked on it
(and its underlying meta-model) throughout all three years of the project.
This long time span led to changes and adaptations of the development
approach over time. As the development of the model is crucial for judging
how much it can be trusted, we describe the development approach and how
we executed it.

\subsubsection{Approach}

Our general approach to developing the base model was to combine
top-down and bottom-up modelling. This means that we started from general quality aspects
and added product factors and measures that are important for these quality
aspects based on our experiences or existing studies. Moreover, we also evaluated 
existing measures, mostly from static analysis tools, investigated their intention and selectively built suitable 
product factors and impacts on quality aspects. We believe both directions are necessary. Modelling
bottom-up is attractive because we know we have measures, most of them automatically collectable,
for the product factors we are building. Yet, to come to a comprehensive assessment of quality aspects,
we need to identify which product factors and measures are missing. Therefore, the combination of top-down
and bottom-up modelling supports the comprehensiveness of the model.

Because we knew that we had to learn while building the base model, we introduced
three large, year-long \emph{iterations}. Our aim was to have one base model after each iteration
that we could further discuss and validate. This included a round 
of feedback from the whole consortium in each iteration, meaning we gave
a base model to people not working on it directly for discussion. In addition,
each iteration ended with detailed reviews by two reviewers from the consortium who were
not directly involved in the base model development. Both the feedback rounds and
the reviews ensured that all experiences of the project members were captured.

The central means for developing the base model and assuring its quality were regular \emph{workshops}.
We used workshops to work on the meta-model as well as on the base model. We conducted
workshops at least at each status meeting (every two to three months) and added additional
workshops as needed. Over time, we used various elicitation and analysis techniques, such as
prototyping, design sketching or pair modelling. A special kind of workshop were tutorial days 
aimed at introducing all members of the consortium to the current state of the base model. This 
also helped us to integrate the experience of all project members.

As we learned that it is easy to build a quality model that cannot be used in a quality
assessment, we introduced the concept of \emph{nightly quality analysis}. This means that
we executed a quality assessment every night on a set of open source projects using the current state
of the base model. Hence, similar to continuous integration, we found problems in the quality
model quickly.

Furthermore, we explicitly \emph{validated} the results of each iteration. We developed four
validation concepts to investigate the structure and contents of the base model as shown in
Table~\ref{tab:validation-concepts}. We wanted to understand the suitability of the structure,
how much the base model assessment results differ between systems, how well the
results fit to expert opinions, and how well practitioners understand and accept the base model
and its results. Each of the validations for every iteration were also reviewed
by two reviewers not directly involved in the validations. The final validations are described in Section~\ref{sec:validation}.

\newcolumntype{R}[1]{>{\raggedleft\arraybackslash}p{#1}}
\renewcommand{\arraystretch}{2}
\begin{table}[htbp]
  \caption{Validation Concepts for the Base Model}
  \centering
  \begin{tabular}{R{2.1cm}p{13.5cm}}
    \hline
    \textbf{Meta-model suitability} & We let experts transcribe (parts of) their existing, diverse quality models into the
    							new structure as given by our meta-model and fill out a questionnaire on how well
							the structure fits their model and needs. Although this does not
							validate the base model directly, the underlying meta-model is a fundamental part
							in terms of the practicality of a quality model. We already published the results of
							this validation in~\cite{Klas.2010}. \\
    \hline
    \textbf{Degree of differentiation} & A quality model that gives a very similar quality grade for all kinds of systems
    							is not helpful. Hence, we ran quality assessments on open source and industrial systems
    							to investigate to what degree the results 
							differentiate the systems. We do not report detailed results in this article.
							Yet, the experiences helped us in the development of the base model
							by understanding what is important (Section~\ref{sec:calibrationbasemodel}).
							Furthermore, we shaped the interpretation model of the quality assessment approach
							(Section~\ref{sec:interpretation}) to support this differentiation. \\
    \hline
    \textbf{Criterion validity} & To understand whether our quality assessment results from the base model matched
    						independent assessments, we compared the assessment results based on the model
						with the independent results for existing systems. First, in case we had a quality ranking of a set of systems or
						sub-systems, we performed a quality assessment using the non-adapted base model
						and compared the resulting quality ranking with the existing one. We did this using expert 
						opinions about open source and industrial systems (Section~\ref{sec:comparison}). Second,
						in case we knew that during certain time periods, changes to a software had been done specifically
						to improve its quality, we assessed the quality at several points in time using the base model and expected to see
						an improvement in the quality grade as well. We performed such a validation with one industrial
						system (Section~\ref{sec:improvement}).\\
    \hline
    \textbf{Acceptance} & Because it is essential that practitioners understand the base model structure,
    					its contents and the assessment results, we conducted assessments of industrial
					systems and inspected the results together with experts for the systems. We
					elicited the experts' understanding and acceptance of the base model and the result of the assessment
					conducted using the base model with a
					questionnaire. We performed these validations for industrial systems (Section~\ref{sec:acceptance}).\\
    \hline
  \end{tabular}
  \label{tab:validation-concepts}
\end{table}
\renewcommand{\arraystretch}{1}

In iteration 1, most of the development of the base model was done by five people 
from one partner organisation. The difficult consolidation and the discussions at the workshops led us to change that approach. We realised that we needed to integrate all partners and set aside dedicated
time slots for working on the base model. Therefore, we introduced the concept of \emph{sprints}.
Each sprint was two weeks long and included four to eight people from two to three partners. We started with 
two days of joint modelling in one location. Afterwards all participants worked independently
on the base model and synchronised their work in daily conference calls. In iteration 2, the sprints were aimed at exploring different areas of what to include in the base model. In iteration 3, we aimed
at consolidating and completing the base model.

\subsubsection{Execution}

In this section, we describe the execution of the approach to building the base model structured along the
iterations. We show an overview of the activities of all three iterations in Table~\ref{tab:iterations}.
We omitted the feedback rounds in the table because we performed one in each iteration to discuss
the current status of the base model. 

\begin{table*}[htbp]
  \caption{Activities for Developing the Base Model}
  \centering
  \begin{tabular}{cp{2cm}p{2cm}p{.9cm}p{3.5cm}p{3.1cm}p{1cm}}
    \hline
    \textbf{It.} & \textbf{Focus} & \textbf{Meta-model workshops} & \textbf{Base model workshops} &
    \textbf{Sprints} & \textbf{Validation} & \textbf{Reviews} \\
    \hline
    1 & Structure and meta-model & 8 & 5 & -- & Meta-model suitability\newline
    									Degree of differentiation\newline
									Criterion validity & 4\\
    2 & Exploring possible contents & 4 & 8 & GUI and accessibility\newline
    									C and maintenance\newline
									Structuredness and portability\newline
									Performance and efficiency & Meta-model suitability\newline
															Degree of differentiation\newline
															Criterion validity & 4\\
   3 & Con\-so\-li\-da\-tion and completion & -- & 3 & Factor consolidation\newline
   									Automation and consistency\newline
									Visualisation\newline
									Calibration and weighting\newline
									Consolidation with specific quality models & Criterion validity\newline
										Acceptance & 4\\
    \hline
  \end{tabular}
  \label{tab:iterations}
\end{table*}

This section and particularly the six steps described below report on the third iteration in more
detail because it had the largest influence on the final contents of the base model.
The main work finalising the base model was done by nine people during a period of three months in which 
we held approx.\ three meetings or conference calls
a week, each lasting about two hours. All in all, this
sums up to approx.\ 80\,hours of joint discussions.
\begin{enumerate}
	\item \emph{Selection of Tools:} In the consortium we decided to
				finalise the base model for the programming languages Java and C\#.
				We chose to use freely available tools as a basis for the measures
				to facilitate the application of the base model. Thus, we selected the
				commonly used tools FindBugs and PMD for Java and Gendarme for C\#.
				Moreover, we included measures available in the quality assessment 
        				framework ConQAT\footnote{\url{http://www.conqat.org/}}. We captured
				the knowledge of our industry partners about important checks in manual 
				reviews as manual measures.
	\item \emph{Distillation of Product Factors:} We used a bottom-up
				approach to define product factors based on the measures collected in step~1. We created an entity hierarchy
				containing concepts of object-oriented programming languages,
				like classes, methods, or statements. Then we assigned each measure
				to a product factor in a series of meetings. If 
				no suitable product factor existed, a new one was created. If
				no suitable entity existed, a new entity was also created. This
				process was guided by the goal of obtaining a consistent and 
				comprehensible base model. Hence, there were extensive discussions
				in the meetings, leading to previously defined product 
				factors being changed. For instance, previously defined product factors were
				split, merged, or renamed.
	\item \emph{Defining the Impacts:} After the product factors had been created in
				step~2, the impacts of each product factor on a quality aspect were
				derived in workshops as in the earlier iterations. Each impact was documented
				in the quality model with a justification how the product factor influences the
				quality aspect.
	\item \emph{Reviewing and Consolidating:} In the next step, the model --
				now consisting of instruments/measures, product factors, impacts, and
				quality aspects -- was reviewed and consolidated. To validate the 
				chain from measures to product factors to impacts, the reviewers asked 
				the following question for each product factor: ``Would the impacts 
				assigned to a product factor be valid if they were directly assigned to each measure of that factor?'' Asking this question meant 
				evaluating whether the chosen granularity of the product factor was right.
				If this question could not be answered positively for a product factor, 
				it had to be split into two or more factors, or a measure
				had to be reassigned to another product factor. In addition, we added the
				top-down view by asking ourselves whether the quality aspects were appropriately
				covered by the impacting product factors and whether the product factors were
				comprehensively measured by the assigned measures.
	\item \emph{Deriving Evaluations:} The creation of all model elements 
				concerned with measurement and the evaluation of the measurement results
				is described in detail in \prettyref{sec:calibrationbasemodel}. 
	\item \emph{Assessment of Coverage:} To assess the completeness of the
				developed base model, we introduced an attribute \emph{Coverage}
				to measure the evaluations. For each measure evaluation, we manually 
				assessed to which extent the available measures covered the product factor.
				The values for \emph{coverage} were determined in workshops and added
				to the model.
\end{enumerate}

\subsection{Contents}

The Quamoco base model -- available together with  
tool support~(see~\prettyref{sec:tool_support}) and in a Web version\footnote{\url{http://www.quamoco.de/}}
 -- is a comprehensive selection of factors and measures 
relevant for software quality assessment. In total, it comprises 
92 entities 
and 
284 factors. 
Since some factors are used for structuring purposes rather 
than for quality assessment, only 
233 factors have evaluations assigned to them. 
Of these, 
201 factors define impacts on other factors, leading to a 
total of 490 impacts. 
These numbers differ from the numbers in our 
previous publication in~\cite{wagner:icse12}
because the model was consolidated at the conclusion of the
Quamoco project, which took place after the publication of that paper.
The relationship between ISO/IEC~25010 characteristics and base model
elements is presented in \prettyref{tab:basemodel-elements}. Note that
product factors can have impacts on multiple quality aspects, and
measures can contribute to multiple product factors, which is why
the sum over a table column can be greater than the total number of
distinct elements of that type. Also, tool-based measures may be
implemented differently for C\# than for Java, so there may be two
tool-based \emph{instruments} for the same measure.

\renewcommand{\arraystretch}{1}
\begin{table}[htp]
  \caption{ISO characteristics vs. base model elements}
  \begin{center}
    \begin{tabular}{lccc}
      \hline
             & \textbf{Product} & \textbf{Tool-Based} & \textbf{Manual} \\
             & \textbf{Factors} & \textbf{Instruments} & \textbf{Instruments} \\
      \hline
			\textbf{Maintainability}        & 146 & 403 & 8 \\
			\textbf{Functional Suitability} &  87 & 271 & 1 \\
			\textbf{Reliability}            &  69 & 218 & 0 \\
			\textbf{Performance Efficiency} &  59 & 165 & 1 \\
			\textbf{Security}               &  17 &  52 & 0 \\
			\textbf{Portability}            &  11 &  20 & 0 \\
      \textbf{Compatibility}          &   0 &   0 & 0 \\
			\textbf{Usability}              &   0 &   0 & 0 \\
			\hline
		\end{tabular}
  \end{center}
  \label{tab:basemodel-elements}
\end{table}

Since the model provides operationalisation for different programming 
languages (cf. Section~\ref{sec:base_model-modularization}), it contains considerably 
more measures than factors: In total, there are 
194 measured factors and 
526 measures in the model. 
For these measures, the model contains 
544 instruments, 
which are divided into 8 manual ones 
and 
536 
that are provided by one of 
12 different tools. 
The tools most relied upon are FindBugs (Java, 
361 rules 
modelled) and Gendarme (C\#, 
146 rules). 
Other tools integrated into our model 
include PMD and several clone 
detection, size, and comment analyses, which are part of ConQAT.

The validity of metrics for measuring and evaluating the presence or absence
of quality characteristics (\eg maintainability) in a software product is well studied but controversial.
While some studies (\eg \cite{Basili.1996},~\cite{Briand.2000},~\cite{Dandashi.2002}) conclude
that software metrics could be used as indicators of quality,
others (e.g.,~\cite{Alshayeb.2003},~\cite{Benlarbi.2000},~\cite{Shatnawi.2010}) deny the validity of metrics.
However, since we put the emphasis on rules of static analysis and adhering to best practices,
we ascribe less importance to the validity of other metrics for our models.

Rules of static code analysis can be used for ascertaining
specific potential flaws within source code, \eg to detect
bug patterns often observed~\cite{ Hovemeyer.2004} or to check for established best practices.
Thus, associating violations of both (due to such rules or best practices such as naming conventions) with factors can make it easier to understand their impact.
No empirical study known to us is currently available, however, that provides statements
about the validity of rule assignments to certain factors. In the specific quality model
for embedded software systems (see section~\ref{sec:specific-eqm}) we evaluated whether
the associated rules and best practices are appropriate and sufficient for certain aspects. 

In the following, we present example product factors including their
respective measures and impacts to illustrate the contents
of the base model. 
An important design decision was to use the
product quality characteristics of ISO/IEC~25010 to describe quality
aspects, because this is the best-known way of describing
software quality. These characteristics all refer to the whole software 
product and are hence modelled in such a way that each characteristic 
refines the factor \emph{quality} with the entity \emph{product} and each
sub-characteristic refines its respective characteristic. 

\subsubsection{Rules of Static Code Analysis Tools}
\label{sec:rules-static}

As described above, the largest fraction of measures refers to static
code analysis tools. One example is the FindBugs rule  
FE\_TEST\_IF\_EQUAL\_TO\_NOT\_A\_\-NUM\-BER, which scans Java code
for equality checks of floating point values with the \emph{Double.NaN} constant.
The Java language semantics defines that nothing ever equals \emph{NaN},
not even \emph{NaN} itself, so that \emph{(x == Double.NaN)} is always false. 
To check whether a value is not a number, the programmer has to call \emph{Double.isNaN(x)}.
This rule is an instrument for the \emph{doomed test for equality to NaN}
measure, which measures the product factor \emph{general expression applicability}
for \emph{comparison expressions}, along with several other measures.
This product factor in turn impacts \emph{functional correctness}, because the developer
intended to check a number for \emph{NaN} but the actual code does not.
It furthermore impacts \emph{analysability}, because understanding the intention of
this construct demands additional effort.

Using automatic source code analysis techniques for quality assessment
implies the risk of false positives, \ie findings reported by the analysis tools
that do not reflect actual problems in the source code. While false positives have
been reported to cause acceptance problems \cite{Bessey2010}, providers of modern 
static analysis tools continuously work on reducing the number of false positives
produced by their tools (\eg \cite{Ayewah2007,Hovemeyer2007}). Another aspect is 
that the calibration (see \prettyref{sec:calibrationbasemodel}) was performed in such a way that normalised measurement results were transformed into utility values based on the value range they produce. This means that the utility mapping functions already
take false positives into account and therefore false positives will not severely impact the assessment results. 

In the Quamoco approach, we adapted the concept of utility value from Multi-Attribute Utility/Value Theory (MAUT/MAVT) \cite{vincke_1992} where utility\footnote{MAUT assumes that decision-making involves risk. For the sake of simplifying the initial quality assessment method, we assume quality assessment to have a riskless character and use the multi-attribute value theory (MAVT). Yet, we will use the term ``utility'' instead of ``value'' to avoid confusion between the meaning of the term ``value'' in the context of MAUT and its common understanding in the software engineering and measurement domains.} represents the strength of preference a decision maker has among alternatives (in our case alternative software products) regarding a specific decision criterion (in our case a quality aspect). Transformation between normalised measurement results and corresponding utility values is performed with the help of utility functions. Utility maps measured, objective values of a quality aspect onto corresponding utilities.

Additionally, measures that we knew had a comparatively high rate of false positives were assigned lower weights during the modelling workshops (see \prettyref{sec:modeldevelopment}).

\subsubsection{Programming Best Practices}

Rule-based code analysis tools cannot detect every kind of quality problem.
Therefore, the base model also contains product factors based on 
metrics and best practices. For example, identifiers have
been found to be essential for the understandability of source
code. Whether identifiers are used in a concise and consistent
manner can only partly be assessed automatically~\cite{2006_deissenboeckf_naming}.
Therefore, the product factor \emph{conformity to naming convention} for \emph{source
code identifiers} contains both automatic checks performed by
several tools and manual instruments for assessing whether identifiers
are used in a consistent and meaningful way.

Another well-known property related to software quality is code cloning.
Source code containing large numbers of clones has been shown to be
hard to understand and to maintain~\cite{Juergens.2009}.
The concept of code cloning is represented in the product factor 
\emph{duplication of source code}, which has negative impacts
on both \emph{analysability} and \emph{modifiability}. It is measured by
\emph{clone coverage} as well as \emph{cloning overhead}. 
Both these
measures have corresponding instruments for Java and C\#, which are
obtained by ConQAT's clone detection functionality.

\subsubsection{Guidelines}

To show that our concepts are not only applicable to 
static analysis of source code, we modelled a subset of W3C's
\emph{Web Content Accessibility Guidelines} (WCAG) 2.0\footnote{
\url{http://www.w3.org/TR/WCAG}}, which is a guideline for
making web content accessible to people with disabilities.
The WCAG differs from the examples described above in several
ways: Most of the contents are phrased as success criteria,
outlining the best way to achieve accessibility and providing
technology-independent instructions to test them. In addition,
the entities of interest are UI components rather than source
code elements. Due to the size of the guideline and the small 
fraction represented in our model, we decided to separate 
this \emph{GUI} module from the base model and provide it separately 
as an example rather than a complete model.

\subsection{Modular Structure} 
\label{sec:base_model-modularization}

\begin{figure}
	\begin{center}
		\includegraphics[width=.6\columnwidth]{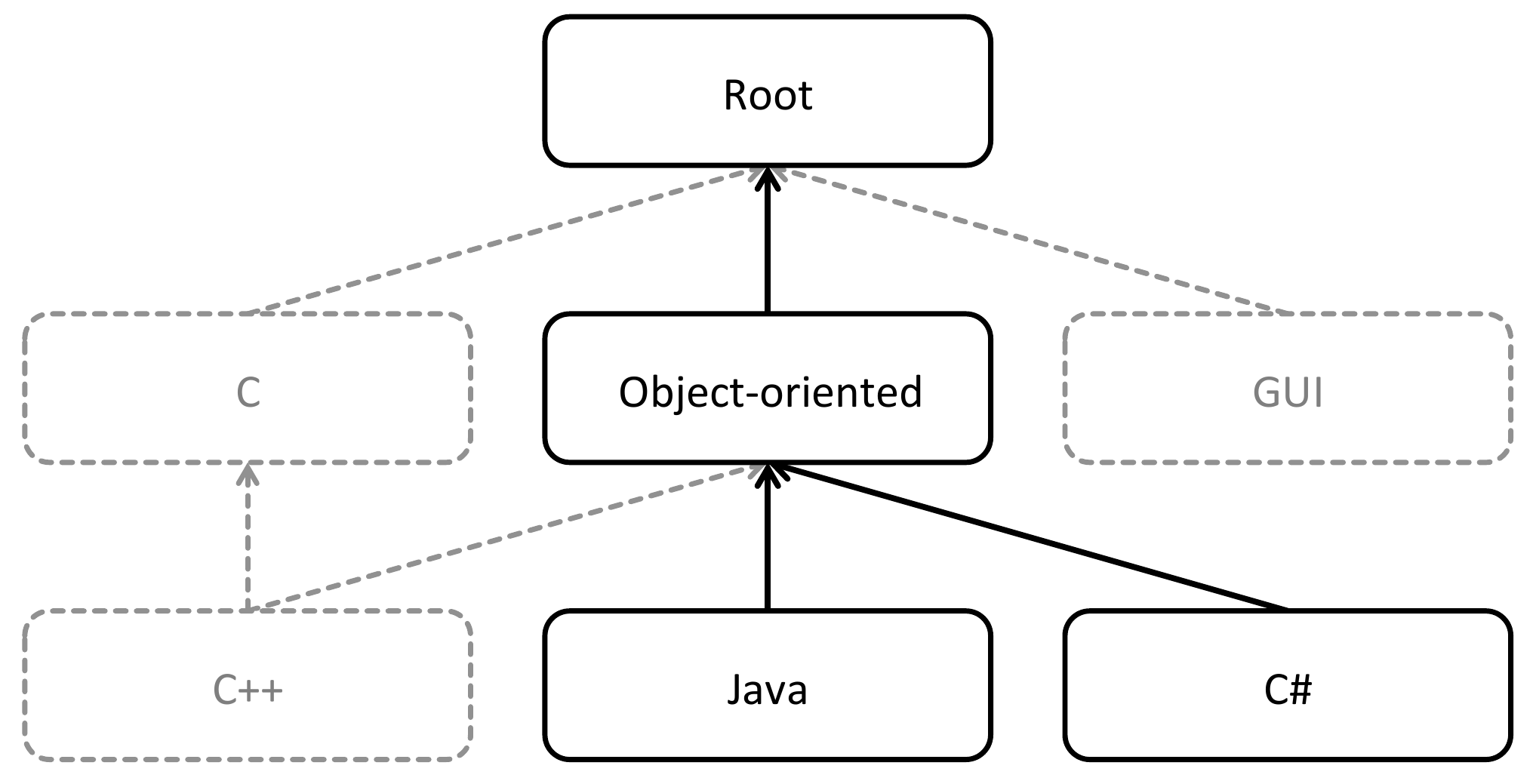} 
		\caption{Modules of the Base Model}
		\label{fig:basemodel:modularization}
	\end{center}
\end{figure}

Modularisation is not only helpful by differentiating between the
base model and specific models but also within the base model itself. 
It allows us to separate more abstract, context-independent elements
from concrete, technology-specific elements.
In Figure~\ref{fig:basemodel:modularization}, 
the modules of the base model are depicted with black lines, while 
experimental modules are depicted in grey. 
In the base model, the module \emph{root} contains the definitions 
of quality aspects as well as product factors that are independent of
programming paradigms or languages. For these product factors, the root module
also contains evaluations as well as impacts on quality aspects, ensuring a
large amount of reusability across the other modules.
We introduced our own module for each programming language in the quality model. 
An intermediate module \emph{object-oriented} defines common concepts of 
object-oriented programming languages (Java, C\#) such as classes or inheritance.

We also used the modularisation concept to integrate specific analysis tools for particular
programming languages. In the module \emph{object-oriented}, we defined a large number of
general measures without connections to concrete tools (e.\,g.,\ number of classes). 
The module for Java then defines a tool for measuring the number of classes in Java systems. This 
way, we support separation between general concepts and specific instruments.

The explicit definition of modules provides several benefits: First, it enables
quality modellers to separately and independently work on modules for different 
technologies and domains.
Second, it allows us to explicitly model the commonalities and differences between several 
programming languages. The module \emph{object-oriented} defines 64 common factors to
be reused in the modules \emph{Java} and \emph{C\#}, which add only 1 and 8 
language-specific product factors, respectively. 

Furthermore, modularisation made it possible to create other modules for conducting 
various experiments without impairing the more stable contents of the base model.
We created two modules for \emph{C} and \emph{C++}, which are not mature yet. Furthermore, we tested
the modelling of non-source code factors in the module \emph{GUI} for graphical user interfaces.

\section{Quality Assessment Approach}
\label{sec:quality_assessment}

A quality model specifies quality in terms of relevant properties of software artefacts and associated 
measures. Yet, to support the assessment of product quality, the quality model needs to be associated 
with an approach to synthesise and interpret the measurement data collected for the product. In this section, 
we specify a quality assessment method applicable for Quamoco quality models. We assume for the
method that we have already adapted the quality model suitably for our context (see Section~\ref{sec:adaptation} for more detail on adapting quality models).

\subsection{Quality Assessment Method}
\label{sec:qualityassessmentmethod}
In this section, we propose a systematic method for assessing the quality of software products. The method relies on the requirements we identified in a systematic literature review and a small industrial survey~\cite{Trendowicz_Kopczynska_2014}. These requirements are briefly summarized in Table~\ref{tab:sqa_requirements}.

\begin{table}[htbp]
	\caption{Requirements regarding a quality assessment method (SQA)}
	\centering
	\begin{tabular}{R{4cm}p{12cm}}
		\hline
		R01. Supports group decision-making and multiple viewpoints & SQA should be transparent and understandable for the quality stakeholders. If the assessment provided by the assessment method deviates from what is expected by the quality stakeholders, then they should be able to easily identify and understand the causes of the discrepancies.\\
		\hline
		R02. Comprehensible to quality stakeholders & SQA should be transparent and understandable for the quality stakeholders. If the assessment provided by the assessment method deviates from what is expected by the quality stakeholders, then they should be able to easily identify and understand the causes of the discrepancies.\\
		\hline
		R03. Handles uncertain information & SQA should be applicable to both certain and uncertain information. When uncertain information is provided, the assessment method should also indicate the uncertainty of the assessments it delivers.\\
		\hline
		R04. Correctly comprehends the preferences of the stakeholders & SQA should assess the quality of a software product as it is perceived (and as it would be assessed) by software decision makers whose quality requirements this software product should satisfy.\\
		\hline
		R05. Copes with incomplete information & SQA should be applicable for incomplete input. When applied to incomplete information, the method should provide reasonable outcomes, probably within a certain uncertainty range.\\
		\hline
		R06. Handles interdependencies between elements of the quality model & SQA should handle potential interdependencies between quality attributes (i.e., elements of the underlying quality model). It other words, the quality assessment should explicitly consider how the quality attributes interact with and constrain each other, and how they affect the achievement of other quality attributes.\\
		\hline
		R07. Combines compensatory and non-compensatory approach & SQA should support assessment for both comparable and non-comparable quality attributes. For comparable quality attributes, the method should support mutual compensation of their "negative" and "positive" values.\\
		\hline
		R08. Supported by an automatic tool & SQA should be supported by an automatic software tool. Moreover, configuring and operating the tool should not require much effort. After configuring, the tool should allow fully automatic quality assessment.\\
		\hline
		R09. Allows for incorporating subjective expert evaluations & SQA should allow for manual assessment of the selected quality attributes and for incorporating these intermediate assessments into the overall quality assessment process.\\
		\hline
		R10. Supports benchmarking & SQA should support comparing software products directly, using their quality assessment results. For this purpose, quality assessment results need to be comparable against each other for different software products, particularly for different versions of the same product.\\
		\hline
		R11. Provides repeatable results (assessments) & SQA should provide repeatable assessments when applied with the same set of alternatives.\\
		\hline
		R12. Custom-tailorable & SQA should allow for assessing quality regardless of the particular structure of the quality problem and the context of the quality assessment. The structure of a quality problem typically refers to the structure of the underlying quality model (e.g., flat list of effort factors or hierarchical structure of aspects and sub-aspects).\\
		\hline
		R13. Supports hierarchical quality model & SQA should operate on a hierarchically structured quality model. In other words, the method should be applicable to quality aspects that are organized in a hierarchical structure.\\
		\hline
		R14. Scalable and extensible & SQA should allow for assessing quality independent of the size and complexity of the underlying quality model (the number of quality attributes and their dependencies). In other words, the applicability of the method should not depend on the size and complexity of the underlying quality assessment problem. Moreover, the assessment method should also allow changing the size and complexity of the underlying quality model, that is, it should not be fixed for a specific quality model.\\
		\hline
		R15. Supports intermediate assessments & SQA should allow for assessing quality at intermediate levels of the hierarchical structure of the underlying quality model. In other words, it should be possible to assess quality aspects at any level of the quality model hierarchy instead of doing a single assessment for the root quality aspect.\\
		\hline
	\end{tabular}
	\label{tab:sqa_requirements}
\end{table}

The quality assessment procedure comprises four basic steps: measurement, evaluation, aggregation, and interpretation. \textit{Measurement} consists of the collection of measurement data for the factors specified at the lowest level of the quality model's hierarchy according to the measures defined in the quality model. \textit{Evaluation} comprises the determination of factor utility values for individual factors based on the collected measurement data. \textit{Aggregation} comprises the synthesis of utilities obtained on individual factors into the total utility of a product under assessment. Finally, \textit{interpretation} is the translation of the relatively abstract utility values into a quality assessment value that is intuitive for human decision makers.
Before a quality assessment can be executed, it first needs to be operationalised. The measurement step may require defining additional measures to ensure that measurement data collected for the same measure across different products are comparable. The evaluation step requires defining utility functions to model the preferences of decision makers with respect to the measures defined for the factors. Aggregation requires defining the aggregation operator to synthesise the utilities of the individual factors across the quality model hierarchy into the total utility  value. This includes assigning factors with numerical weights to reflect their relative importance to a decision maker during the aggregation. Finally, interpretation requires defining an interpretation model that will translate utility values into a quality assessment value that is intuitive for a human decision maker. Users of the quality model and the quality assessment method can (and should) perform operationalisation steps prior to quality assessment make the approach fit their specific context (\eg they should adjust utility functions and factor weighting to reflect their specific preference regarding the importance of individual factors).

\begin{figure}[htbp]
	\begin{center}
		\includegraphics[width=.7\columnwidth]{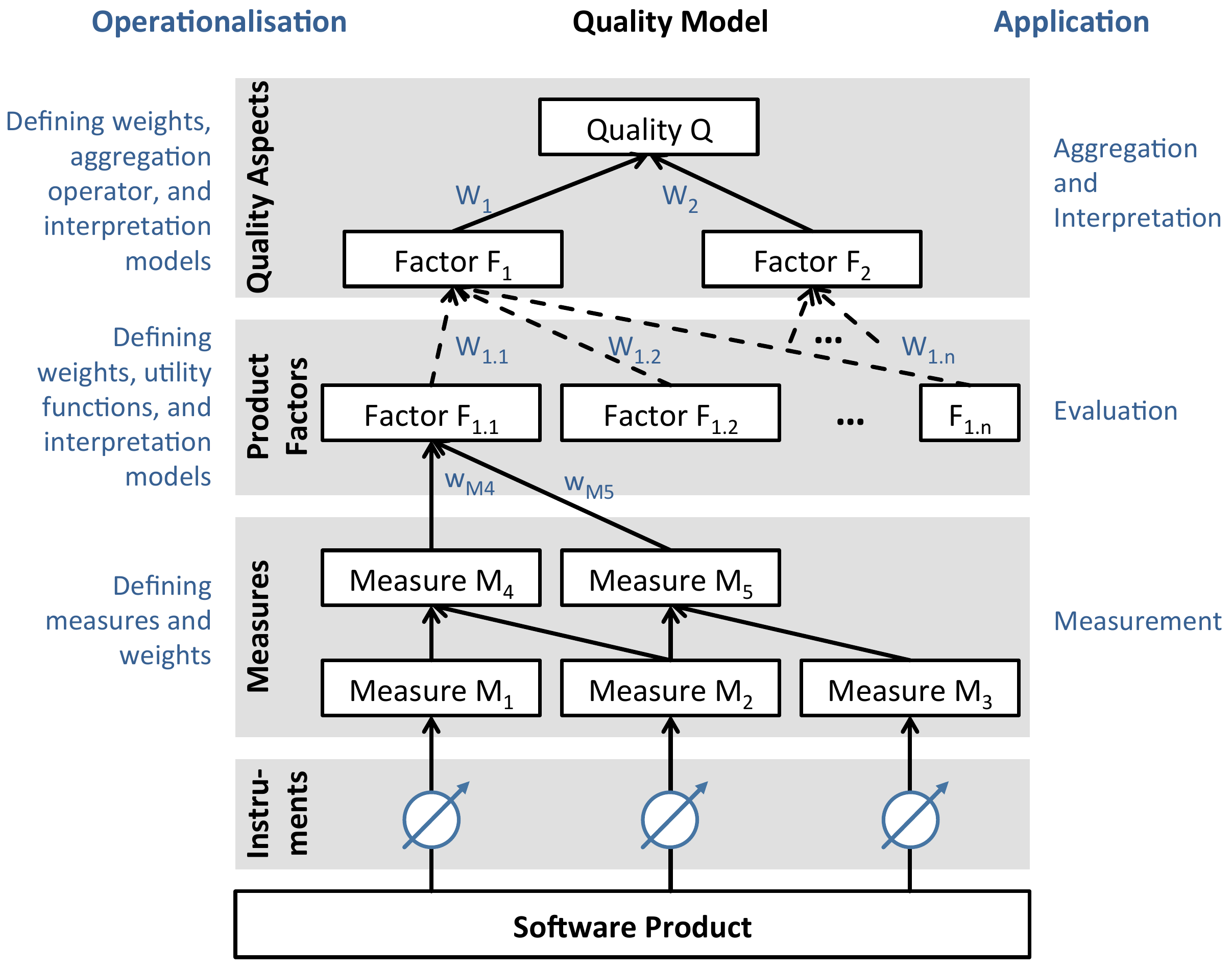}
		\caption{Overview of the Quality Assessment Approach}
		\label{fig:ass1}
	\end{center}
\end{figure}

The Quamoco quality assessment method specifies both operationalisation of the quality model for the purpose of assessment and the quality assessment procedure. Figure~\ref{fig:ass1} shows the basic steps of the model's operationalisation (left side of the figure) and of the quality assessment procedure (right side of the figure) mapped onto the related elements of the quality model. These basic steps and their outputs correspond to the generic process of Multicriteria Decision Analysis (MCDA) (\eg \cite{Dodgson2000}). 

\begin{figure}[htbp]
	\begin{center}
		\includegraphics[width=.7\columnwidth]{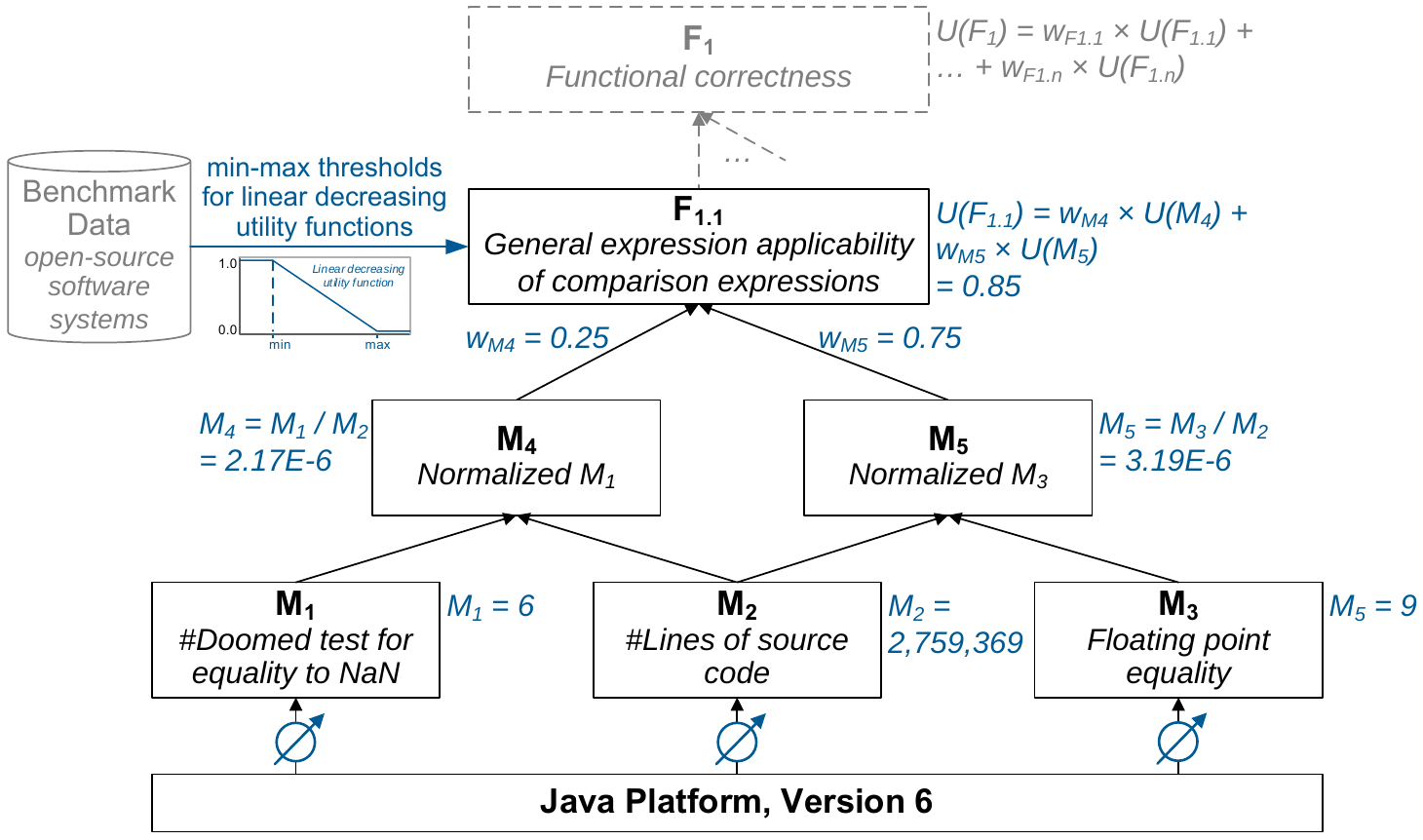} 
		\caption{Example Quality Assessment}
		\label{fig:ass5}
	\end{center}
\end{figure}

In the remainder of this section, we will discuss the basic operationalisation and assessment steps and illustrate them with an excerpt of the base model described in Section~\ref{sec:rules-static}. Figure~\ref{fig:ass5} presents the example quality model with a summary of the assessment results.

\subsubsection{Measurement} 
\label{sec:defining:measures}
We start the operationalisation by normalising the measures associated with the product factors. Normalisation is aimed at ensuring that the measurement data collected for the same measure are comparable across different products. For this purpose, we define \emph{normalisation measures}. It has to be noted that cross-product comparability of the results may be negatively affected
by the selected normalisation measure. For instance, the measure LOC will have different effects for products using different programming languages. Hence, we employ more abstract measures, such as \emph{number of classes}, for normalising class-related measures, where appropriate.

In the application, we collect values for the measures and normalisation measures using the associated instruments, which can be manual
or automatic. This step is well assisted by our tool support (Section~\ref{sec:assessment-engine}) by running automated analysers, 
transforming all values into a uniform format, and normalising them. 

\paragraph{Example} In the operationalisation step, we normalise the base measures which count the number of specific rule violations. We normalise the base measures \textit{$M_1$: Doomed test for equality to NaN} and \textit{$M_3$: Floating point equality} by the base 
measure \textit{$M_2$: Lines of code} into the derived measures $M_4$ and $M_5$, respectively. Through normalisation, these measurements of $M_1$ and $M_3$ become comparable between software systems of different sizes. For the source code of the Java platform, version 6, we obtain $M_1 = 6$, $M_2 = 2,759,369$, and $M_3 = 9$. Consequently $M_4 = 2.17 \times 10^{-6}$ and $M_5 = 3.26 \times 10^{-6}$.

\subsubsection{Evaluation} 
\label{sec:defining:utility:functions}
In the operationalisation step, we specify the evaluations for factors (Section~\ref{sec:meta-model}). We define a utility function for each measure of a product factor at the lowest level of the quality model. These functions specify the utility each measure has in the context of the product factor with which it is associated. To ensure that the evaluation will be understandable, we use only simple linear increasing and decreasing functions with two thresholds \emph{min} and \emph{max}, which determine when the factor is associated with the minimal (0) and maximal utility (1). 

Once we have decided on the type of function (decreasing or increasing), we determine the thresholds for the function using a benchmarking approach. The basic principle of benchmarking is to collect a measure for a (large) number of products (benchmarking base) and compare the measure's value for the product under assessment to these values. This allows us to decide if the product has better, equally good or worse quality than other products. The details of how we determine the thresholds are described in the appendix. 

During the application, we calculate the defined evaluations using the collected measurement data. Hence, we evaluate the utility of all product factors at the lowest level of the quality model. 

A common requirement not addressed by most existing quality assessment approaches is how to cope with incomplete measurement data~\cite{10.1109/QSIC.2006.13}. In case of incomplete inputs to the assessment, we assess quality using best-case and worst-case values for the missing data and express the uncertainty using the resulting range. 

\begin{figure}[htbp]
	\begin{center}
		\includegraphics[width=.6\columnwidth]{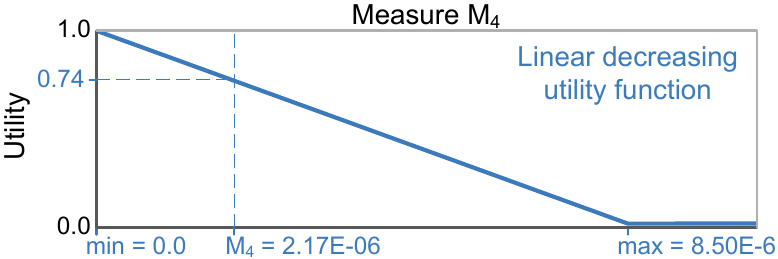} 
		\caption{Linear Utility Function}
		\label{fig:ass2}
	\end{center}
\end{figure}

\paragraph{Example} To operationalise the evaluation step, we had to define utility functions for the measures $M_4$ and $M_5$ which are directly connected to the factor $F((1.1)$. The higher the value of each of these measures and the worse it is for software quality, the lower should be the associated utility. To reflect this, we selected simple decreasing linear utility functions. We derived $min$ and $max$ thresholds for the utility functions based on the benchmark measurement data from 120 open-source projects. The derived thresholds were $min(M_4) = 0, max(M_4) = 2.17 \times 10^{-6}$ and $min(M_5) = 0, max(M_5) = 3.26 \times 10^{-6}$. (Figure~\ref{fig:ass2}) illustrates the utility function for the measure $M_4$.

The utility function can then be used for deriving the utilities with which the measures $M_4$ and $M_5$ contribute to the factor $F_{1.1}$. We calculate $U(M_4) = 0.74$ for measure $M_4$ (see Fig.~\ref{fig:ass2}) and $U(M_5) = = 0.89$.

\subsubsection{Aggregation} 
During operationalisation, we assign numerical weights to the elements of the quality model, specifically to all quality aspects and product factors, and additionally to the measures that are assigned to the product factors defined at the lowest level of the quality model hierarchy. Numerical weights represent the relative importance of the adjacent elements of the quality model to the decision makers. 

We suggest forming relevance rankings based on available data or expert opinion. We can then use the \emph{Rank-Order Centroid} method~\cite{Barron1996} to automatically calculate the weights from the relevance ranking according to the \textit{Swing} approach~\cite{Edwards1994}. 

During application, we use the weights in the bottom-up synthesis of factor utilities along the hierarchy of the quality model. For this purpose, we need an appropriate aggregation operator. We use a weighted sum operator as an easily understandable and relatively robust aggregation approach.

\paragraph{Example} In the operationalisation step the aggregation operator is selected and appropriate importance weights are assigned to factors and quality aspects in the quality model. In our example, we used simple weighted sum aggregation.  Weights assigned to factors (and quality aspects) quantify how important the factor is in the quality model relative to its sibling factors. That is, factors that have the same direct parent in the quality model hierarchy. The importance of the i-th factor relative to its sibling factors is measured as the cardinal weight $w_i$ such that: $w_i \in [0, 1]$ and the weight across all sibling factors (including the i-th factor) sums up to 1. In our example $M_4$ was rated as three times less important for $F_{1.1}$: \emph{General expression applicability of comparison expressions} than the second measure $M_5$. The calculated importance weights are thus $w_{M_4} = 0.25$ and $w_{M_5} = 0.75$. 

The aggregated utility of factor $F_{1.1}$ is calculated as follows: $U(F_{1.1}) = w_{M_4} \times U(M_4) + w_{M_5} \times U(M_5) = 0.25 
\times 0.74 + 0.75 \times 0.89 = 0.85$. The same aggregation principle would be applied for higher levels of the quality model, \eg for $F_1$: \emph{Functional correctness} the aggregated utility would be $U(F_1) = w_{1.1} \times U(F_{1.1}) +  ... + w_{1.n} \times U(F_{1.n}) = 0.02 
\times 0.85 + ... = 0.82$. Yet, for the sake of simplicity, we limit our example to the level of factor $F_{1.1}$.

\subsubsection{Interpretation}
\label{sec:interpretation}
The activities described in the previous sections support the decision maker in interpreting a factor's utility, for example, if it is high or low. The objective of interpretation is to map the ratio-scale utility onto a more intuitive scale, for instance onto an ordinal scale of school grades or traffic lights. During operationalisation, we define the form of the interpretation model. 

\begin{figure}[htbp]
\begin{center}
\includegraphics[width=.6\columnwidth]{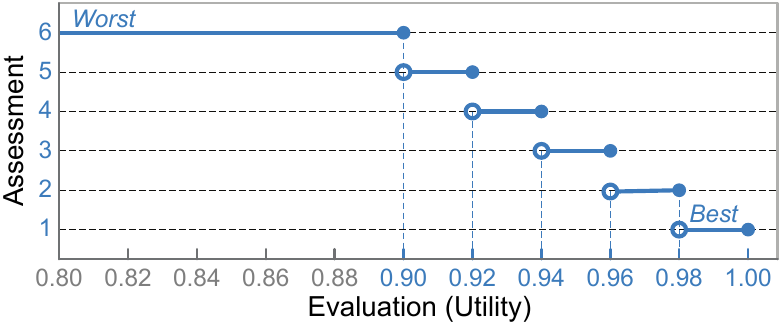} 
\caption{Interpretation Model}
\label{fig:ass4}
\end{center}
\end{figure}

Figure \ref{fig:ass4} illustrates our proposed interpretation model for which we chose the metaphor of school grades. The thresholds used in the model correspond to the coefficients used in German schools to convert students' performance scores assigned for spelling tests. More than 10\% incorrectly spelled words disqualify the test with the worst grade of 6. We adopt this scale for the purpose of quality assessment by decreasing the 
quality grades by one every 2\% until the best possible grade of 1 is reached.

This step should be further supported by visualisations that prevents the assessor from getting lost in the detailed quality model and its assessment results. For example, a sunburst diagram enables a focused investigation across several hierarchy layers, or a Kiviat diagram helps to compare the results for different systems or subsystems.

\paragraph{Example} In the operationalisation step, we select the interpretation model that will be used for converting relatively incomprehensible utility values into a quality assessment understandable for human experts in the specific context of quality assessment. In our example, we selected a German school grades model which maps the percentage of task fulfilment ranging between 0\% and 100\% onto a 6-point grade scale. The advantage of this model is that the source scale 0-100\% corresponds to the utility scale 0-1 and school grades are intuitive for every human expert (considering country-specific grading systems).

Following the interpretation model in school grades (Figure \ref{fig:ass4}), a utility of 0.82 for \emph{F1.1: Functional correctness} gives a grade of 6 (worst).

\subsection{Operationalisation of the Base Model}
\label{sec:calibrationbasemodel}

We performed the entire operationalisation of the assessment steps already for the base model; so it
is readily available for assessments. This was done as part of the base model development 
(Section~\ref{sec:modeldevelopment}). The operationalisation may still need to be adapted for specific 
contexts but can provide a quick idea of a system's quality.

For the measurement step, two of our experts reviewed each measure to determine an appropriate
normalisation measure based on a defined set of rules. 

In the evaluation step, we performed benchmarking on open source systems. 
We calibrated the C\# part of the base model with 23~open source systems, which
we manually prepared for the quality assessment. For the Java module of the base model, 
we used the repository SDS~\cite{Bajracharya:2009lr} as a benchmarking base. 

According to \cite{Gruber2010}, in an optimal benchmark base, not only the 
programming language is important for the selection of proper benchmarking projects, but the 
size and application domain should also be considered. The SDS repository 
contains about 18,000 open-source Java projects. These projects were mostly retrieved 
from open-source databases such as \emph{Sourceforge} via a web-crawling 
approach. In essence, this repository contains mirrors of the version control 
repositories. Thus, these repositories usually contain not only the most recent version
of a software but also \emph{branches} and \emph{tags}. 

For benchmarking, we prefer using only one version of each software project, and hence we decided to use the
most recent one. For finding the most recent one, we used a heuristic that is 
able to handle the most commonly used directory structures for Java projects. If
there is a directory called ``trunk'', then this is used; otherwise the root directory
is used. The heuristic is also able to handle a directory structure where the root 
directory contains multiple Java projects in single directories, each of them 
containing a directory named ``trunk''. Given that these are the most commonly
used directory structures for Java projects, we conclude that the heuristic should 
work in most cases. Moreover, we manually verified it for a dozen projects.

The SDS repository only contains the source code, but not the binaries. For certain 
quality measurements (\eg FindBugs), however, binaries compiled with the debug option of the Java compiler are 
needed. We compiled all systems in a batch approach because the effort for manually 
configuring and compiling them is prohibitive. The compilation of all systems took about
30~hours, executed in parallel on 12~PCs. Of all available systems, 
about 6,000 were compiled successfully. Others could not be compiled because of missing 
external libraries or because code needed to be generated during the build process. 

To get a representative set of systems from the remaining 6,000 systems, we 
randomly selected systems of different size classes. We know the distribution of systems 
in different size classes from~\cite{Wicks.2005} and selected systems accordingly: 
39~systems were larger than 100\,kLoC, 42~systems were between 10~and 100\,kLoC, and 
19~systems were between 5~and 10\,kLoC. Since the SDS repository contained no systems
larger than 250\,kLoC, we included 10~systems of this size class that were available
within the consortium. Most of the used systems were open-source systems and, hence, 
might not be representative of commercial software. Yet, we do not see any possibility to perform
such an approach with commercial systems because we would not be able to get a reasonable
number. Furthermore, with open systems, the analysis can be easily repeated and replicated.

We automatically calculated the thresholds for the linear distributions using the approach described in
the appendix. Finally, two of our experts reviewed these thresholds
for each measure by benchmarking them together with supporting descriptive statistics 
for plausibility.

In the aggregation step, we had to define weights. For the quality aspects, we extracted the relative 
importance of the quality aspects from the results of a survey done by us~\cite{Wagner2010} that had more than
100 responses. For example, functional suitability and reliability were considered very important, while 
portability and accessibility were considered less important. For all other elements -- \ie measures 
assigned to product factors and impacts targeting quality aspects -- we used our regular workshops
and sprints to determine relevance rankings. 

In the interpretation step, we used the model described above based on school grades.
To investigate the interpretation model's discriminative power (IEEE~1061) for assessments with the 
base model, we also exploited the results from these 120 systems. As we randomly selected systems 
from a large number of systems, we expected wide differentiation in the quality assessment results. 
The interpretation model fulfilled this 
expectation in that the results have a wide spread: The assessments distributed the sample systems 
across the entire range of the interpretation scale (grades 1--6) with only few systems being assessed at 
the extreme ends of the scale (very good or very bad) and a slight tendency towards good assessment results 
(1: 9\%, 2: 37\%, 3: 31\%, 4: 13\%, 5: 6\%, 6: 4\%). The distribution of the results also fits well with the typical 
distribution expected for German school tests and is therefore in concordance with the selected interpretation metaphor.


\section{Tool Support}\label{sec:tool_support} 

We developed a comprehensive tool chain for the
application of the Quamoco approach in software projects. The tooling supports
building and editing quality models, adapting a quality model to organisation- or
project-specific needs, assessing software systems according to a
quality model,  and, finally, visualising the results of a
quality assessment. The tool support presented here is available from
the Quamoco website\footnote{\url{http://www.quamoco.de/tools}}.

\subsection{Quality Model Editor} 

The quality model editor is built on the Eclipse Platform and the
Eclipse Modeling Framework. It allows editing quality models that
conform to the Quamoco meta quality model. To support the
modularisation concept, each module of a
quality model is stored in a separate file. The content of the model
can be navigated via different tree views, which allow form-based
editing of the attributes of model elements.

Validation during editing helps the modeller create models that adhere
to meta-model constraints, consistency rules, and modelling best
practices. A simple validation rule checks for unreferenced model
elements. A more sophisticated rule ensures that for model elements
referenced in other modules, an appropriate \emph{requires} dependency
between the modules is defined. The editor employs the Eclipse marker
mechanism for displaying error and warning messages in a list and
provides navigation to affected elements. The user is further assisted
by an online help feature that displays context-sensitive help content
depending on the current selection in the editor. The help texts
explain the concepts of the meta quality model and contain a guideline
with best practices for quality modelling.

The editor also supports several steps of the assessment method 
(Section~\ref{sec:qualityassessmentmethod}). All operationalisation activities 
take place in the quality model editor, such as the
definition of measures, utility functions or weights. Furthermore, it allows
running the assessment engine (Section~\ref{sec:assessment-engine})
on a given quality model and load the results for visualisation
(Section~\ref{sec:visualisation}).

\subsection{Adaptation Support}
\label{sec:adaptation}
 
Since quality depends on the context, \eg the distinctive characteristics of a domain or technology, a quality model needs to be adapted to its environment. The meta-model and the modularisation concept provide the necessary foundations separating contents for different contexts. To support efficient and consistent adaptations beyond modules, we additionally provide an adaptation method~\cite{2011_klaes_QM_adaptation} and tooling.

The adaptation approach provides a set of rules and automation based on the premise that all quality models used and produced have their structure defined by the meta quality model. It can be used to adapt quality models at different abstraction levels (such as general-use models, organisation-level models, or project-level models). The adaptation approach explicitly considers the goal and context of the model used to perform initial adaptation; subsequently, adaptation tasks are generated to obtain a consistent and complete model.

The steps for adapting a quality model are: 
\begin{enumerate}
\item Define the goal of the resulting model based on the organisation/project context and its software quality needs
\item Identify an existing quality model that best matches the target model's goal defined in the previous step
\item Perform automated pre-tailoring by eliminating all unneeded elements (all elements that do not satisfy the model's goal)
\item Adapt the model in an iterative manner until it satisfies the identified goal which implies adjusting the remaining elements and, where needed, adding new elements to the model
\end{enumerate}

The adaptation approach of Quamoco is implemented in a plugin for the model editor called the \emph{adaptation wizard}, which consists of a series of dialogs that guide the user through the whole adaptation of a quality model. The first step consists of specifying the goal of the model meaning the type of artefacts that the model should be applied to, the perspective (\eg management or customer), the quality focus (\ie the key quality aspects that should be addressed in the model) and its context (\eg paradigm or programming language).

Using this goal, the adaptation wizard searches the workspace for models that match the specified goal. The goals of existing models are extracted from their content. For example, the existing factors are used for matching the quality focus. The context is extracted from the modules and from text tags that can be given to any of the elements. Text tags may include many different things describing the context, such as domain (\eg railway, medical devices, embedded systems, information systems), methodologies, practices, or technologies supported (\eg component-based software development, agile development, open source software, custom development, C++, Java and automatic measurement tools). Text tags are assigned manually. Many elements can be tagged simultaneously. Per default, the base model is available for adaptation. When the base model is selected for adaptation, all its elements are automatically tagged in accordance with the modules they belong to. Other general-use models, organisation-level models, or project-level models can be added to the workspace by manually importing them using the structure provided by the meta-model. 

The person performing the adaptation is guided by the tool in choosing the best matching model and selecting relevant elements. Irrelevant elements are eliminated automatically. For example, if a product factor is not selected as relevant, it is deleted, as are its impact-relationships to any quality aspect. The wizard can be configured to execute the process top-down, beginning with the selection of factors or entities, or bottom-up, beginning with the selection of measures.

Once irrelevant elements have been removed, the wizard uses the previous actions and the model's goal to generate a list of further adjustments needed to satisfy the model's goal and scope. The required adjustments are presented as a list of adaptation tasks, which includes reviewing, adding, and modifying specific model elements (\emph{adaptation tasks view}). These tasks need to be performed manually, as they involve some decision making from the person performing the quality model adaptation. For example, if a measure is added, it needs to be marked as a normalisation measure or a factor needs to be associated with it.

All changes are documented and shown in the adaptation history view, which also lists the justifications for changes that are automatically generated and can be extended by the user. The adaptation history is a good basis for reviewing and judging the modifications performed and can thus serve as input for future quality model certification activities.

\subsection{Quality Assessment Engine} 
\label{sec:assessment-engine}

The quality assessment engine automates the application activities
of the assessment method (Section~\ref{sec:qualityassessmentmethod}).
It is built on top of the quality
assessment toolkit ConQAT which allows creating quality dashboards by
integrating various quality metrics and state-of-the-art static code
analysis tools.

The connection between quality modelling and assessment is achieved through
the automated generation of a ConQAT analysis configuration from a
quality model. For the assessment of a software system, the quality
assessment engine is provided with the adapted quality model, the source code of the
software system to be assessed, the generated ConQAT configuration, and
manual measurement results stored in an Excel file. This
allows extending the tooling with custom analyses needed for assessments
based on extensions to the base model.

For reasons of convenience, the editor supports direct execution of the assessment 
for the base model by specifying the location of the
software system to be analysed. The output of the quality assessment
engine has two formats. The first is an HTML report which allows
inspecting the results of an assessment from within a browser, thus
not requiring the tooling and the quality model. The second output
format is a result data file, which can be imported into the editor
for interactive inspection of the results. To detect decays
in quality as early as possible, the quality assessment engine can
also be run in batch mode, which allows us to include it into a continuous
integration environment.

\subsection{Visualisation} 
\label{sec:visualisation}

Once the result data file produced by the quality assessment engine has been
imported into the quality model editor, a number of visualisations
support a detailed analysis of the assessment, for example, for
tracking quality issues from abstract quality aspects to
concrete measures.

\begin{figure}[htb]
\centering\includegraphics[width=.6\columnwidth]{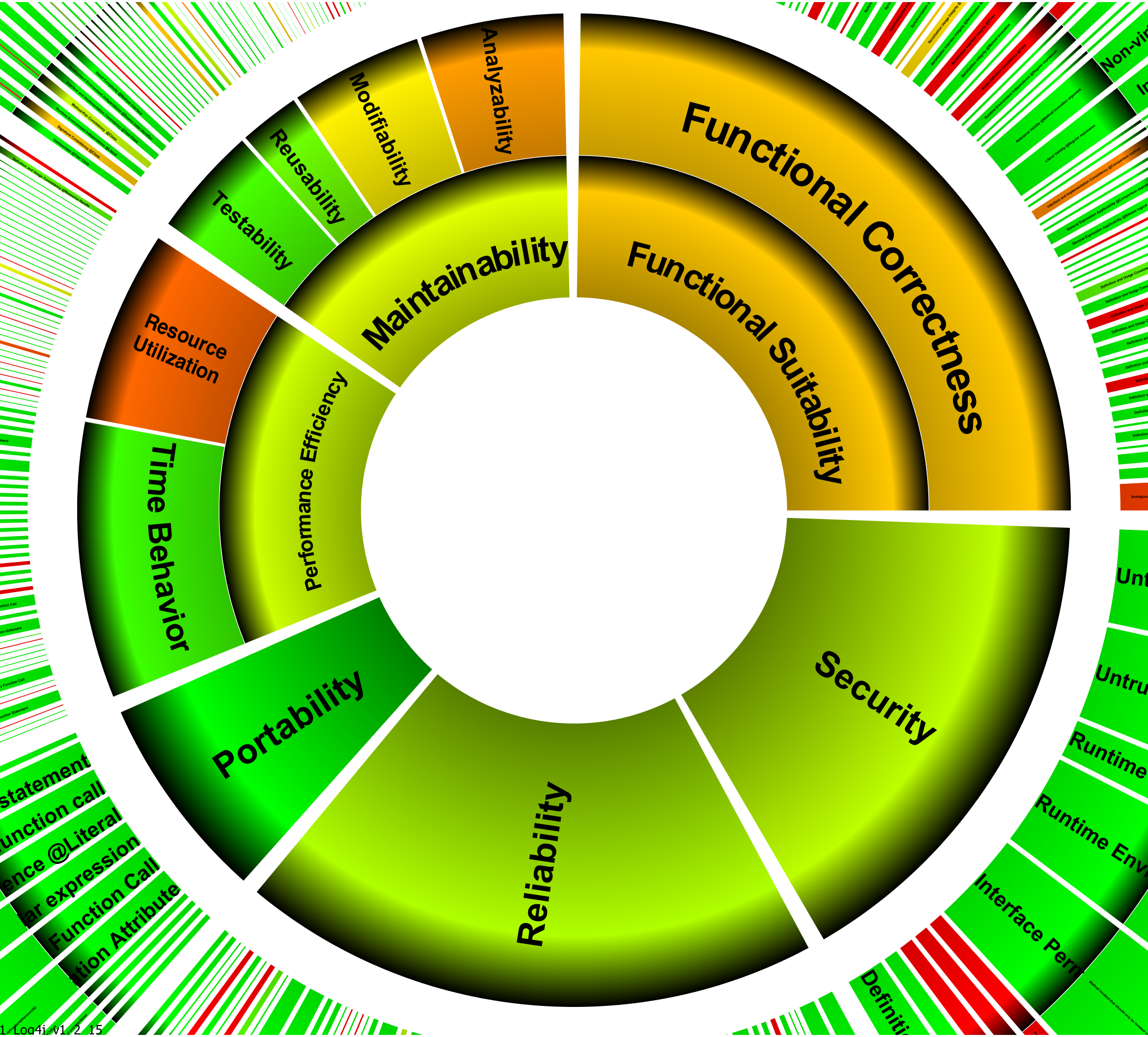}
\caption{Sunburst Visualisation}
\label{fig:sunburst}
\end{figure}

For example, a sunburst visualisation (Figure \ref{fig:sunburst}) of
the results provides an overview of the assessed factors and their
impacts by showing the complete hierarchy of the model all at once. Each
segment in the chart corresponds to a factor in the quality model
with the angle of a segment denoting the factor's importance. The
hierarchy is given by the segment's adjacent positioning from the
centre outwards. The colour of a segment visualises the assessment
result for the assessed system. The colour range from green via yellow to red indicates good, average, or bad results. To improve comprehensibility, it is
possible to zoom into a factor, which maximises the space occupied by
that factor in the visualisation.

\begin{figure}[htb]
\centering\includegraphics[width=.6\columnwidth]{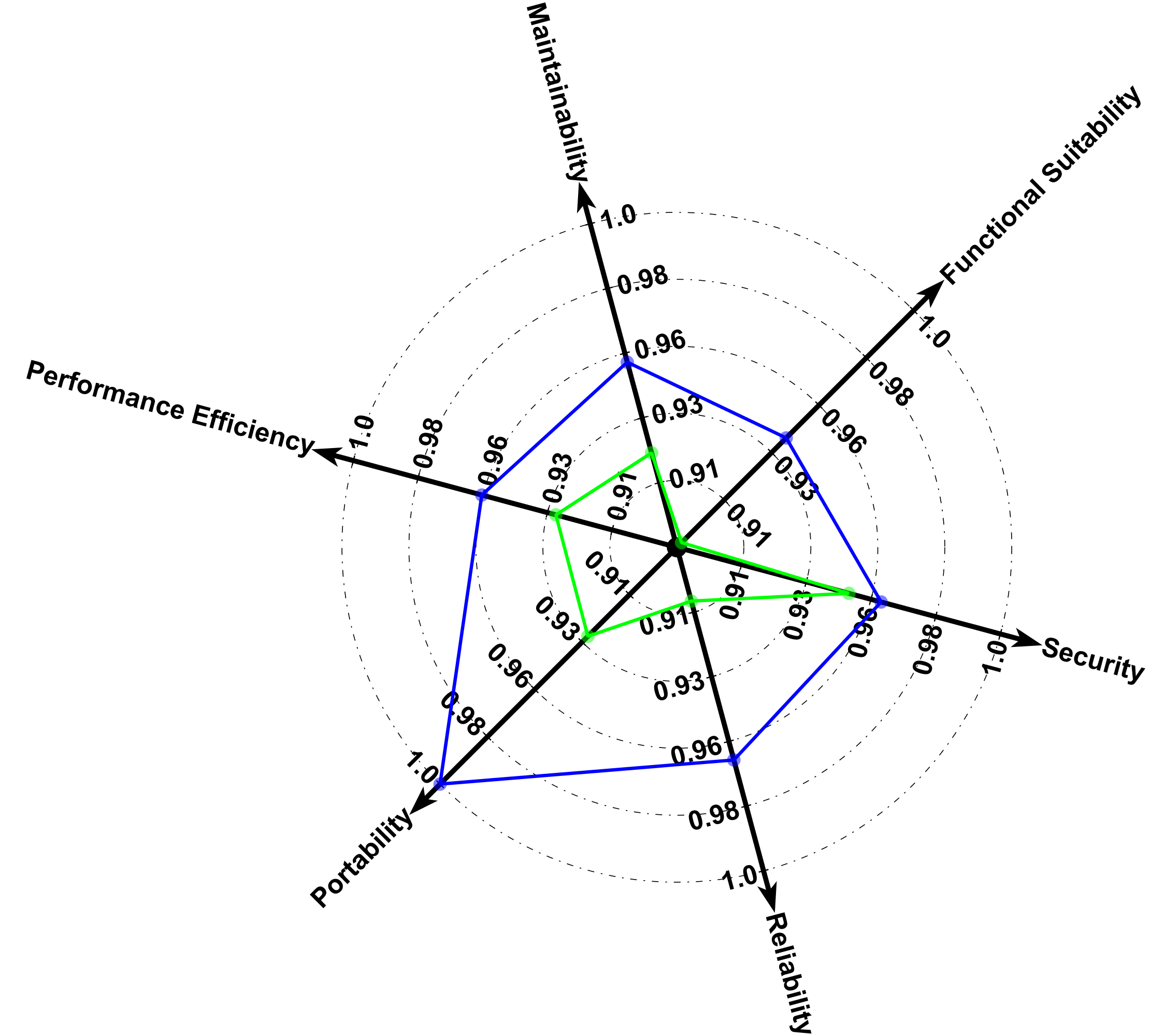}
\caption{Kiviat Diagram}
\label{fig:kiviat}
\end{figure}

A Kiviat diagram (Figure \ref{fig:kiviat}) offers an alternative
high-level visualisation of the assessment results. It specifically aims
at supporting the comparison of quality assessments for multiple
systems (or multiple versions of the same system). The spokes in the
diagram correspond to the top-level factors in the quality model and
the coloured data points correspond to the factor assessment results
for the systems. This visualisation allows quickly spotting those
quality aspects where the assessed systems differ.

\section{Empirical Validation}
\label{sec:validation}

This section presents the empirical validation of the base 
model. We focus on the Java model because only this model was thoroughly calibrated, as 
explained in \prettyref{sec:calibrationbasemodel}. We 
consider the calibration of the C\# model as weak because we had only a dataset of~23 systems; thus, we
exclude this part of the model from the empirical validation.

In the following subsections, we investigate the Java base model and its quality assessments 
using three research questions:
\par\noindent\textbf{RQ 1}: Does the base model provide valid assessment results regarding the quality of software systems?
\par\noindent\textbf{RQ 2}: Can the base model be used to detect quality improvements over time in a software system?
\par\noindent\textbf{RQ 3}: Is the approach of assessing quality using the base model accepted by practitioners?

Research questions RQ~1 and RQ~2 address the usefulness of the Quamoco approach from a technical point of view: They check whether the Quamoco approach provides valid assessment results in the context of typical application scenarios. RQ~3 focuses on the usefulness and applicability of the approach as perceived by practitioners.

We limit the evaluation scope of RQ~1 and RQ~2 to overall quality assessment results. In practice, a statement on overall quality is important, for example, to identify software products or releases with low quality (cf. RQ1) or to get an early warning if the quality of a product in the company portfolio decreases. It can also be used to control the success of quality improvement initiatives (cf. RQ2).

In each subsection, we first present the study goal and then the research design chosen for the study; next, we describe the execution of the study and its results. Finally, each section closes with a discussion of the results and the most relevant threats to validity. The raw data for all the validations are available upon request. Where possible, we made it available in \cite{wagner:quamoco-data}.

\subsection{Comparison of Software Products}
\label{sec:comparison}

To answer RQ~1, we have to evaluate whether the base model provides valid assessment results, meaning that the assessment results are in concordance with the results obtained by another independent and valid approach for assessing product quality. Checking criterion validity is also a common approach in other scientific disciplines (e.g., psychology) when introducing new measurement instruments to quantify an abstract concept such as quality.  

\subsubsection{Design} To evaluate the validity of the quality assessments, we need an independently obtained criterion for product quality that we can compare with our assessment results. Since no measurement data were available that directly measure quality for a set of software products that we could assess using the base model, we utilised expert-based quality judgments as the independent criterion. 

For the comparison of different software products, we used the quality rating obtained in the \emph{Linzer Software-Verkostung} \cite{Gruber2010} for a set of five open source Java products. The \emph{Linzer Software-Verkostung} was directed and carried out by a member of the Quamoco team before the start of the Quamoco project. Due to the involvement of a member of the Quamoco team, we had access to all detail data gathered during the experiment. The rating in the \emph{Linzer Software-Verkostung} is a ranking of the five Java systems based on a combination of ratings provided independently by nine experienced Java experts. 

For a serious evaluation of Java source code it is necessary for each expert to have not only experience in software development or software quality management, but to also be familiar with the programming language itself. During the selection process for the \emph{Verkostung}, we paid attention to the experience, background and environment of potential experts. Because each expert needs time to provide a reasonable and independent quality evaluation for each software product, only five systems could be considered as data points in the comparison. Although, from a scientific point of view, it would be preferable to have more data points to increase the probability of obtaining statistically significant results, from a practical point of view this could not be realised in the presented study. In the end, nine experts were accepted for the \emph{Linzer Software-Verkostung}. Each participant had at least seven years of experience in software development. 

\emph{Evaluation criterion and hypotheses:} To measure validity and ensure comparability with other studies, we used the validity criteria proposed in the IEEE standard~1061 for validating software quality metrics. The standard proposes a set of criteria, but most of them assume that the collected measures and the independent criterion both use an interval or ratio scale; only one can also be applied to ordinal scale data. In our case, the results of the base model assessments were provided as a value characterising the product quality between grade 1 (best possible) and grade 6 (worst possible) on a ratio scale and the assessment results of the expert judgements were provided on an ordinal scale as a ranking from best (rank 1) to worst (rank 5) product. Consequently, we had to limit our investigation to the validity criterion \emph{consistency} (cf.\ IEEE~1061) which can be applied to ordinal-scale data. In our case, it characterises the concordance between the product ranking based on the assessment results provided by our model and the ranking provided independently by a group of experts. This means that we determined whether the base model could accurately rank the set of assessed products/subsystems with respect to their quality (as perceived by experts) and is thus valide regarding this external validity criterion.

Following the suggestion of IEEE~1061, we measured \emph{consistency} by computing the Spearman's rank correlation coefficient (r) between both rankings where a high positive correlation means high consistency between the two rankings. Since we wanted to check whether a potentially observed positive correlation was just due to chance or was a result of using an appropriate quality model, we stated hypothesis $H_{1_A}$ (and the corresponding null hypothesis $H_{1_0}$). We tested both with the confidence level 0.95 ($\alpha$ = 0.05):

$H_{1_A}$: There is a positive correlation between the ranking of the systems provided by the base model (BM) and the ranking of the systems provided by the experts during the Linzer Software-Verkostung (LSV).

\begin{eqnarray*}
H_{1_A}: r (\mbox{rankingBM}, \mbox{rankingLSV} ) > 0 \\   
\mathrm{\ie} H_{1_0}: r (\mbox{rankingBM}, \mbox{rankingLSV}) \le 0 \\
\end{eqnarray*}

\subsubsection{Execution} 
We used the Java part of the base model to assess the quality of five open source products for which independent expert-based assessment results of the Linzer Software-Verkostung were available: JabRef, TV-Browser, RSSOwl, Log4j, and Checkstyle. In this context, it was necessary to collect the manual measures defined in the base quality model for the selected open source products. We used the measurement specifications provided by the base model and collected measurement data for all five products. The collected measurement data was independently reviewed before assessing the products. We ordered the assessed products by the results for their overall quality provided by the base model and compared them with the ranking provided by the Linzer Software-Verkostung. 

The nine experts of the Linzer Software-Verkostung were guided by a questionnaire that forced them to concentrate on the quality attributes analysability, craftsmanship, reliability, and structuredness -- thus a combination of maintainability and reliability topics. Each of these quality attributes was explicitly defined in the questionnaire. Additionally, we provided an ordinal scale for evaluation. For this ordinal scale, we gave an explanation as to the conditions under which to use which rating. 

\subsubsection{Results} 
Table \ref{tab:verkostung} shows the resulting grade and product ranking for the base model assessment as well as the ranking from the Linzer Software-Verkostung. Although the base model did not provide exactly the same ranking as the experts, the calculated Spearman's rho correlation is r = 0.9, which is close to a perfect correlation of 1. The reason is that the two systems with different ranking results have ranking results close to each other -- only their places are switched. Hypothesis $H_{1_A}$ can also be accepted with a p-value of 0.05, meaning that there is a significant positive correlation between the ranking provided by the base model and the ranking provided by the Linzer Software-Verkostung. 

\renewcommand{\arraystretch}{1}
\begin{table}[htp]
\caption{Comparison of the Base Model Assessment Results and the Results of the ``Linzer Software-Verkostung''}
\begin{center}
\begin{tabular}{lrrrr}
\hline
\textbf{Product} & \textbf{LOC} & \textbf{Grade BM} &	\textbf{Rank BM} & \textbf{Rank LSV}\\
\hline
Checkstyle &	 57,213 & 1.87	& 1 &	1\\
RSSOwl &		 82,258 & 3.14	& 2 &	3\\
Log4j &			 30,676 & 3.36	& 3 &	2\\
TV-Browser &	125,874 & 4.02	& 4 &	4\\
JabRef &		 96,749 & 5.47	& 5 &	5\\
\hline
\end{tabular}
\end{center}
\label{tab:verkostung}
\end{table}%

\subsubsection{Interpretation} 
The assessments of the overall product quality for the five investigated systems turned out to be largely consistent and thus valid when compared to an independent criterion for quality, in this case provided in the form of an expert-based assessment. It is important to understand that the assessment results are based on the overall calculated results for all measures and product factors. We did not explicitly test whether the same results could be obtained with a smaller number of measures or product factors, as in our understanding the factors and measures we consider in our quality model represent real quality requirements regardless of the numerical impact on the evaluation results. 

\subsubsection{Threats to Validity} 
\textit {Conclusion Validity.} The generalisability of our results is limited by the fact that the scope of the empirical validation was limited to five medium-sized open source systems written in Java. 

\textit {Internal Validity.} We cannot fully guarantee that the criterion chosen for the validation, namely the expert-based quality rating, adequately represents the quality of the products. Nevertheless, if we consider the ranking of the projects by the experts and calculate Kendall's W concordance coefficient, we get a value of 0.6635, meaning that there is a strong concordance of the rankings of the individual experts with a significance level of $\alpha = 0.01$. We therefore can safely assume that the quality rating of the experts reflects the common understanding of internal software quality.   

\textit {Construct Validity.} In the Linzer Software-Verkostung the experts were guided by a questionnaire that contained the quality attributes to be considered in the expert-ranking. To reduce ambiguities or misunderstandings, the quality attributes were explicitly defined. 

\textit {External Validity.} The experts of the Linzer Software-Verkostung were carefully selected, i.e.\ only experts with significant experience (7+ years of development experience in industry) were part of the experimental setting and therefore manifest a high level of quality awareness. The study was based on open-source software and did not consider industrial strength software. This is a threat to validity as we know from other work~\cite{Mayr.2014} that there are, for example, significant differences in internal software quality between safety-critical and non-safety-critical software. We therefore cannot conclude whether our approach would also hold for safety critical systems.  

\subsection{Sensing Quality Improvements}
\label{sec:improvement}

In a second study, we knew from the quality experts of the project that they had invested effort into enhancing the maintainability of the 
system. In this study from the automation domain (steel production), we analysed four versions of the Java software using our base model. The major goal was to validate whether our base model would reflect the assumed quality improvements claimed by the quality managers (RQ~2). Thus, the validation should also show to some extent whether our base model reflects the common understanding of quality (in the context of maintenance) of experienced developers. 

\subsubsection {Design}
To carry out this study, we performed an ex-post analysis of several versions of a software product, where the responsible technical product manager stated that the project team explicitly improved quality starting right after the rollout of version 2.0.1. The quality improvements were not directed by static code analysis tools at all but by the experience of the software developers. 

\subsubsection {Execution}
We used the base model to assess the quality of four versions of the software product, including all four versions between 2.0.1 and 2.2.1, in our analysis. We presented the results to the project team and asked for their feedback by means of an interview. As the project team had no concerns about the results, further investigations were not necessary. 

\subsubsection{Results} Table \ref{tab:quality_improvement} shows the results of the assessment using the base model. The first column shows the version number of the software, the second column shows the corresponding calculated quality grade.

\begin{table}[htp]
\caption{Quality Improvements in an Automation Software Project}
\begin{center}
\begin{tabular}{cc}
\hline
\textbf{Version} & \textbf{Grade BM}\\ 
\hline
2.0.1 & 3.63\\
2.0.2 & 3.42\\
2.1.0 & 3.27\\
2.2.1 & 3.17\\
\hline
\end{tabular}
\end{center}
\label{tab:quality_improvement}
\end{table}%

The results show steady improvements (as expected) for versions 2.0.2, 2.1.0 and 2.2.1. Our 
assessments reveal an improvement of 0.46 from version 2.0.1 to version 2.2.1. 

\subsubsection{Interpretation}
The developers relied on their understanding of software quality during the improvement. As
the assessments using the base model also show an increase in quality, this is an indicator 
that the model reflects the quality understanding of experienced software engineers.

\subsubsection{Threats to Validity} 
\textit{Conclusion Validity.} The generalisability of our results is limited by the fact that we only had one project, where the improvements were carried out without the use of static code analysis tools but rather relied on the experience of the developers. 

\textit{Internal Validity.} A possible internal threat would be that the improvement activities were influenced by the base model or the findings of static analysis tools. We mitigated that risk by ensuring that the software developers of the project were neither involved in the development of the base model nor did they use static code analysis for their improvements.

\textit{Construct Validity.} There is no threat to construct validity, as we analysed the project ex-post, i.e.\ there was no relation between the improvement actions and our analysis.

\textit{External Validity.} The external validity of the results is limited by the fact that we had only  access to one industry project where the software developers explicitly invested in quality without using static code analysis tools.

\subsection{Acceptance by Practitioners} 
\label{sec:acceptance}

In this third study, we aimed to investigate the acceptance of the base model by 
practitioners, investigating \emph{perceived suitability} and \emph{technology acceptance}. We asked experts to assess a software system that was part of their direct work environment using
the base model and interviewed them subsequently about their experience.

\subsubsection {Design} 
We created a questionnaire for semi-structured interviews. To operationalise our study goal, we broke down the concepts to be characterised into sub-concepts. \emph{Perceived suitability} contains four sub-concepts: \emph {Transparency of assessments} is defined as the quality of the base model that enables the user to easily understand an assessment. \emph{Comprehensibility of quality definition} is defined as the quality of the base model parts to be arranged logically and consistently. \emph {Consistency of results with own perception} means that when the model is applied for the same software products, the quality assessments using the base model coincide with the assessments provided by experts. Finally, \emph {overall suitability judgment} means that the base model meets the requirements for measuring and assessing quality. \emph{Technology acceptance} groups \emph {perceived usefulness} and \emph {perceived ease of use} using a shortened version of the definition of Davis~\cite{1989_davis_tam}. \emph{Comparison with ISO/IEC} groups the sub-concepts \emph {ease of adoption} and \emph {definition of quality}.

A detailed operationalisation is shown in Table~\ref{tab:drill-down}. For closed questions, a 5-point rating scale was used: -2: strongly disagree, -1: disagree, 0: neither agree nor disagree, 1: agree, and 2: strongly agree. After answering each closed question, the experts could give additional comments to extend their answers. 

For all variables, we had the following hypotheses: $H_0: \mathit{Median} \le 0 $ and $H_1: \mathit{Median} > 0$. With this we were able to test whether the experts significantly agreed with our statements. 

\subsubsection {Execution} \label{sec:validation-acceptance-execution}
We performed eight interviews. The participants had between 4 and 20 years of experience in software development; only one of them explicitly mentioned having experience in quality management. We chose the participants opportunistically based on personal contacts in the companies. Each interview was conducted separately and independently from the other interviews following four steps: introduction and training, quality model editor walk-through, evaluation result drill down, and closing. 

During introduction and training (5--10 minutes), the interviewer briefly explained the Quamoco goals and status, the expert introduced him-/herself and his/her role, and the interviewer recorded the experience level. Then, the interviewer gave an overview of the interview process and an example of how the approach and the tool work.

During the model editor walkthrough (10--15 minutes), the interviewer explained the model's high-level structure. Then the interviewer drilled down into examples and explained low-level elements. 

The third part of the interviews consisted of asking the expert to assess, from his/her point of view, a system that was part of his/her direct work environment. The interviewer presented the base model assessment results of the system starting top-down with the overall quality result and drilling down into the factors of special interest to the expert (25--30 minutes).

The closing (5 minutes) consisted of letting the expert answer the questions regarding technology acceptance and practicability in comparison to ISO/IEC and make additional comments. We chose
the quality model in ISO/IEC~25010 because it is based on the well-known 9126 and, therefore, provides us with a comparatively solid comparison point. If we were unable to improve beyond the standard in some way,
it would not be advisable for practitioners to adopt our approach.

\subsubsection {Results} The hypotheses were tested with one-sample Wilcoxon signed-ranks tests of location, with $\alpha$ = 0.05. Table~\ref{tab:drill-down} shows the median, the test results\footnote{Bold values are significant at 0.05.}, and the median absolute deviation (MAD) as a measure of variability in the provided answers.
 
\begin {table}
\caption{Results of Interviews (Closed Questions)}
\label{tab:drill-down}
\begin {tabular}{ p{10cm}  c  c  c}
\hline
\textbf{Items} & \textbf{Median} & \textbf{p} & \textbf{MAD}\\
\hline 
\\
\textbf {Perceived suitability}\\

\\\emph {Comprehensibility of quality definition}\\
Q1: The number of impacts associated with the currently assessed ISO/IEC~25000 factor is acceptable. & 1 & 0.128 & 0.5\\

Q2: I agree with the justification of the relationship (impacts) from product hierarchy factors into ISO/IEC~25000 factors. & 1 & \textbf {0.017} & 0.0\\

Q3: The number of measures associated with the impact factors of the currently assessed ISO/IEC 25000 factor is acceptable. & 1 & 0.128 & 1.0\\

Q4: I agree that the measures actually measure the currently assessed impact factors. & 1 & \textbf {0.023} & 0.0\\

\\\emph {Transparency of assessments} \\

Q5: The calculation steps that lead to an assessment are transparent. & 1 & \textbf {0.003} & 0.0\\

Q6: The assessment steps performed (\eg aggregation, ranking) are familiar. & 1 & \textbf {0.017} & 0.0\\

Q7: I understand the assessment steps and could explain them to someone else. & 1 & \textbf {0.007} & 0.0\\

\\\emph {Consistency with own perception}\\
Q8: I consider the assessment results provided by the base model valid, according to my opinion about the quality of the product. & 0 & 0.510 & 0.5\\

\\\emph {Overall suitability judgement} \\
Q9: I think this approach is suitable for measuring and evaluating software quality. & 1 & \textbf {0.004} & 0.0\\

\\\textbf {Comparison with ISO/IEC}\\
\\\emph {Definition of quality} \\
Q10: The base model has a more transparent definition of quality than ISO/IEC 9126/25000. & 2 & \textbf {0.012} & 0.0\\

\\\emph {Ease of adoption}\\
Q11: The base model could more easily be adopted to be applied in practice than ISO/IEC 9126/25000. & 2 & \textbf {0.017} & 0.0\\

\\\textbf {Technology acceptance}\\
\\\emph {Perceived usefulness} \\
Q12: Using the base model in my job would increase my productivity. & 0 & 0.510 & 0.0\\

\\\emph {Perceived ease of use} \\
Q13: I would find the base model easy to use.  & 1 & \textbf {0.018} & 0.5\\

\hline
\end{tabular}
\end {table}

\emph{Perceived suitability}: $H_1$ holds for all sub-aspects of transparency of assessments, \ie assessments can be easily understood. $H_1$ holds for half of the sub-aspects of comprehensibility of the quality definition. $H_1$ does not hold for consistency with own perception; the assessments produced by the base model were considered neither appropriate nor inappropriate. $H_1$ holds for overall suitability, which means that the approach was considered suitable for measuring and evaluating software quality.

\emph{Technology acceptance}: There was significant agreement that the model is easy to use, but there was no agreement about perceived usefulness.

\emph{Comparison with ISO/IEC}: $H_1$ holds for all sub-aspects; in this case the median is 2: the reviewers strongly agreed that the base model could be adopted more easily for application in practice than ISO/IEC 9126/25000 and that the base model has a more transparent definition of quality than ISO/IEC 9126/25000.

Some interviewees gave additional comments, but we could not reasonably analyse them systematically, as they were very few.
Instead, we used them to augment the quantitative results. The participants mentioned that the trustworthiness of third-party measurement 
tools may be a problem as well as the language independence of impacts, the precision of impact justifications, the subjectively determined 
rankings, the understandability of the rankings, the justification of the rankings, and the difficulty to understand how the calibration limits are 
obtained.

Favourable feedback included that the modelled relations are comprehensible and reasonable, that the impact justifications are reasonable and 
are a good idea, that school grades obtained by comparison with reference projects are a good idea, that the unification of different tools is 
good, and that using an ISO/IEC view is a good thing for managers. Moreover, the participants considered the whole approach as the best that 
can be done with static code analysis; as good for getting an overall view on the quality of a software product; as consistent; as an approach 
that clarifies software metrics; and as a good starting point for people who have not used ISO/IEC 9126/25000 before.

\subsubsection {Interpretation} 
Overall, the interviewees followed our expectations and found the base model suitable, better in comparison
to ISO/IEC~25000 and acceptable for their work. A threat in building operationalised and, therefore, large and detailed
models with complex structures is that practitioners will not understand and accept them. We could mostly show the contrary
here. 

The first exception is whether the assessment result is valid. 
Opinion were mostly undecided, meaning that the experts were not sure about the validity. We assume this was most likely 
caused by the time constraints described in \prettyref{sec:validation-acceptance-execution} and the fact that the base
model was not adapted to the experts' specific context. We have to investigate this further (\eg by conducting
a dedicated validity study offering more time). Since the experts were undecided, we conclude that this does not invalidate 
the base model. The second exception is that the interviewees were also undecided as to whether the base model would increase 
their productivity. This also needs further investigation but could be explained by the fact that most interviewees were 
developers who do not regularly perform quality assessments.

Regarding the number of elements in the base model (cf. Q1 and Q3), there was no broad consensus either. This can 
directly be explained by the low number of participants and the high number of elements in the quality model, combined with
the uneven distribution of model elements across quality aspects (see \prettyref{tab:basemodel-elements}). The 
participants answered these questions after assessing different subsets of the base model, so the answers have to
be treated individually and an insignificant overall rating is not a threat to the acceptance of the base model as such.

\subsubsection {Threats to Validity} 
\textit{Conclusion Validity.} Because of the low number of participants, the results have low statistical power.

\textit{Internal Validity.} The interviews were carried out independently of each other in terms of time and space. Nonetheless, we could not control interactions outside the interviews, with the potential consequence that the participants might have known about the interview contents beforehand (Contamination). The participants in the interviews came from industry; consequently, they represent practitioners. We cannot claim that the sample is representative (Selection). The participants did not only specialise in selected quality characteristics but also work as developers, which for some of them is the main role. This may be a threat to validity because their daily tasks involve quality assessment only indirectly. The same applies to the systems assessed, which were different for all participants according to their specific areas of expertise.

\textit{Construct Validity.} Not all items used were validated. This may be a threat to validity because their meaning may be misunderstood (Explication of constructs). To mitigate this threat, domain experts reviewed the operationalisation of the questionnaire. Not all concepts used were measured using multiple items. This may be a threat to validity because some of them may have been measured only partially (Mono-method bias). 

\textit{External Validity.} For the interviews, the base model was shown using the quality model editor, which is not a tool used daily by the experts (Interaction of setting and item). To mitigate this threat, we asked the participants whether they understood each of the views provided by the tool presented during the model editor walkthrough. All participants understood the tool.

\section{Specific Quality Model for Embedded Systems} 
\label{sec:specific-eqm}
One of the important properties of our meta-model (see Section~\ref{sec:meta-model}) is the ability to extend the base model to adapt it to the needs of more specific programming languages or systems. To show this we developed a quality model for embedded systems (ESQM). In this section we present the content of the ESQM and validate it to find out whether the assessment results are in concordance with the results obtained by another independent approach (see Section~\ref{sec:esqm-validation}). Furthermore we discuss (see Section~\ref{sec:esqm-basemodel-comparison}) in more detail which parts of the base model could be reused.  

The specific quality model ESQM enhances the base model with factors 
that are specific for embedded systems (details on the kind of extensions we made to the base model can be found in Section~\ref{sec:esqm-basemodel-comparison}). In~\cite{Mayr.2012} we published the systematic 
development approach for ESQM, which was driven by 
capturing the requirements for embedded systems with a comprehensive literature analysis 
on commonly accepted guidelines and quality standards for embedded and safety-critical systems. In this paper, we put emphasis on the developed model and relate it to the base model to show that extending an existing model can help to ensure consistency and reduce effort for model construction. 
These guidelines and standards are a good source for eliciting specific embedded systems quality 
requirements as they directly provide (among other things) recommendations for embedded systems code quality.

We examined the umbrella standard for functional safety, IEC 61508 Part 3~\cite{IEC61508:2010},
the standard for medical device software IEC 62304~\cite{IEC62304:2006},
as well as the standard for railway applications EN 50128~\cite{EN50128:2001}. 
These standards play a vital role for safety-critical (embedded) systems and have led to generic and programming language independent requirements. Additionally, we considered the MISRA C:2004~\cite{MISRA-C:2004}, MISRA C++:2008~\cite{MISRA-CPP:2008} 
guidelines as well as the JSF AV C++ coding standards \cite{JSF:2005} 
for programming language-specific requirements elicitation. Ultimately, a total of 60 requirements 
emerged from these sources.
Additionally, we provide information on the iterative validation cycles of the model in~\cite{Mayr.2012}.
In this section we outline the resulting ESQM, sketch its validation, and investigate the model elements that are
necessary in addition to the base model. 

\subsection{Contents}

\begin{figure}
  \begin{center}
    \includegraphics[width=.6\columnwidth]{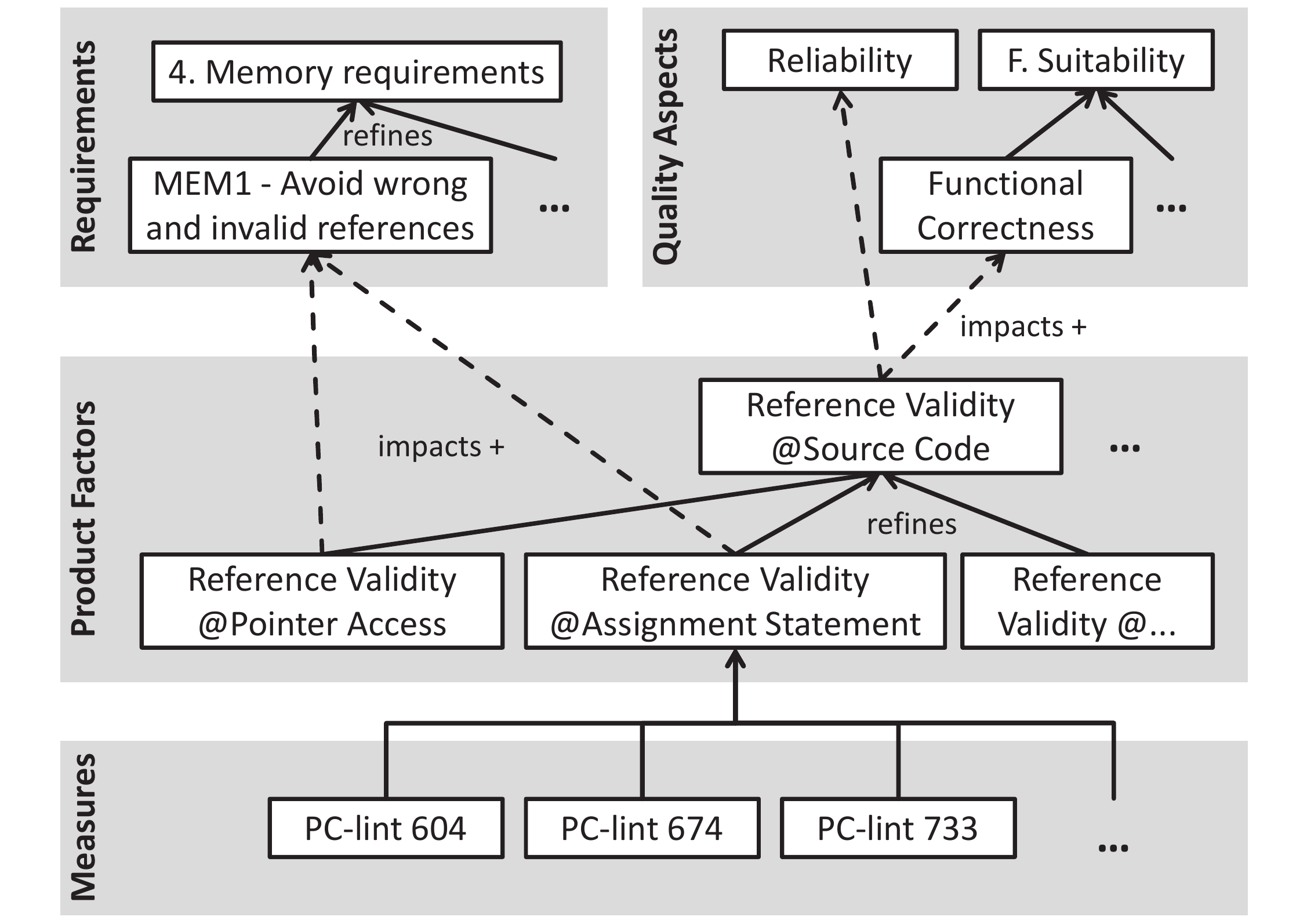} 
    \caption{ESQM Model with Factors (Excerpt)}
    \label{fig:specific-eqm}
  \end{center}
\end{figure}

The excerpt in Figure~\ref{fig:specific-eqm} provides an overview of the ESQM. 
The measures quantify the product factors, which again refine other product factors or 
impact the requirements or the quality aspects.
The model uses the modularisation concept of the meta-model and distinguishes between a 
C module, which is the core module, and a C++ module.

In total, 
we modelled 162 product factors on the leaf level of the product factor hierarchy with 
336 measures.  
Besides systematically constructing product factors and operationalising them with measures, we also defined the
impacts of the product factors on the requirements (177 impact relations) and on the ISO/IEC 25010 quality 
aspects (128 impact relations).

Furthermore, we also applied the assessment method defined in Section~\ref{sec:qualityassessmentmethod} and added
evaluations for the ISO/IEC 25010 quality aspects and for the identified requirements. For the ISO/IEC 25010 quality
aspects, aggregation functions are provided for all quality aspects, which allows us to calculate
an overall quality statement for an embedded software system.

For the calibration of the linear evaluation functions for the measures, we collected normalised measurement data 
from 17 open source projects (8 written in C and 9 written in C++), 
identified potential outliers in the data using box plots,
and took the minimum 
and maximum of the non-outlier values as the 
thresholds for the respective evaluation function.

\subsection{Validation}
\label{sec:esqm-validation}

We applied the validation concept \emph{criterion validity} (see Table~\ref{tab:validation-concepts})
to the ESQM as well to understand whether the assessment results are in concordance with the results 
obtained by another independent approach for assessing product quality. For this purpose, we 
compared the results of the automatic assessment for three industrial software products with the 
judgement of one expert who knows these three products well from a quality perspective. The expert 
is a senior engineer with more than 12 years of experience in embedded software quality. Table~\ref{tab:specific-eqm2}
gives an overview of the results.

\begin{table}[htb]
\caption{Comparison of the Assessment Results and an Expert's Opinion}
\label{tab:specific-eqm2}
\begin{center}
\begin{tabular}{lccc}
\hline
\textbf{Product} & \textbf{Grade ESQM} & \textbf{Grade Expert} \\
\hline
A       &  1.2 & 1 \\
C       &  1.8  & 3\\
B       &  3.1 & 5\\
\hline
\end{tabular}
\end{center}
\end{table}

The total result (using German school grades, where 1 is the best grade and 6 is the worst grade) 
shows a clear order of the quality of the products, with product A being the best product, 
product C being second, and product B being the worst product. The assessment method results of 
ESQM for these three products are similar, i.e., the order of the products is the same as the expert's 
opinion. To make the result clearer, we calculated the grades with higher precision, \ie with 
one decimal place. 

The automatic assessment method calculates better grades, although it keeps the same 
order as the expert. Reasons for this might be that the ESQM quality model focuses on quality 
requirements that are directly related to the particularities of embedded systems and that the model does not consider 
more general quality factors that are independent of a specific application domain. We assume that 
the expert considered these more general aspects in his rating.

\subsection{Comparison with Base Model}
\label{sec:esqm-basemodel-comparison}

ESQM uses all concepts as defined in the meta-model and applied in the base model. In addition, 
we found it straightforward to keep ESQM compatible with the base model's properties and entities used. 
Table~\ref{tab:specific-eqm1} gives an overview of the number of properties and entities of the complete base model (i.e.\ Java and C\#) that we reused and changed in the ESQM.

\begin{table}[htb]
\caption{Reuse of Entities and Properties in ESQM}
\label{tab:specific-eqm1}
\begin{center}
\begin{tabular}{lrc}
\hline
\textbf{Elements}         & \textbf{ESQM} & \textbf{Base Model} \\
\hline
\emph{Entities}         & 87   & 92 \\
~~added          & 30    & - \\
~~split          & 11    & - \\
~~merged         & 17    & - \\
~~removed        & 29   & - \\
~~renamed        & 7    & - \\
\hline
\emph{Properties}       & 32   & 43 \\
~~added          & 6    & - \\
~~split          & 1    & - \\
~~merged         & 4    & - \\
~~removed        & 14   & - \\
~~renamed        & 1    & - \\
\hline
\end{tabular}
\end{center}
\end{table}

For the entities, considerable restructuring was necessary. We added a 
number of entities, as in C/C++ the declaration and definition of variables or methods has to be
distinguished. Furthermore, C++ has some specifics not available in other object-oriented programming
languages, such as comma operators. We split some entities, as the underlying measures sometimes provide more
specific support for those specialised entities (\eg not only for literals in general but also for string literals and 
numeric literals). We removed those entities that do not make sense for C/C++ (\eg interfaces).

Table~\ref{tab:specific-eqm1} shows, however, that many of the properties already defined in the base model could be
reused. Only minor changes were necessary (adding, splitting, merging, renaming) to model the properties
that are typical for embedded systems. We removed 14 properties from the base model. These were mainly properties 
tied to entities that are specific for programming languages like Java and C\# (\eg interfaces) or 
properties that are not related to source code.

\section{Summary \& Conclusions}
\label{sec:conclusion}

A gap exists between abstract quality definitions provided in common 
quality taxonomies, such as ISO/IEC~25010, abstract quality assessment procedures,
such as ISO/IEC~25040, and concrete measurements. Our overall aim 
is to close this gap with the help of operationalised quality models. Because of the domain specificity
of quality and the potential detailedness of such operationalised quality models, we also
aimed at developing modular and adaptable models. To conclude, we summarise 
our contributions, discuss the lessons learnt and draw conclusions and give directions for future work.

\subsection{Summary}

We have shown six contributions for achieving our goal: (1) We developed 
an explicit meta-model that allows us to specify operationalised quality models with 
the flexible but well-defined concepts of \emph{factors}, \emph{impacts} between factors, 
\emph{measures} for operationalising the factors, and modules. (2) Using this meta-model, we built a 
broad, largely technology-independent base model, which we operationalised 
for the programming languages Java and C\#. The freely available and extendable 
base model captures the most important statically measurable product factors and their impacts on product 
quality as defined in ISO/IEC~25010. (3) We provided a quality assessment approach
and a corresponding operationalisation of the base model, which enables us to use it 
for transparent and repeatable quality assessments. 

(4) We evaluated three aspects 
of the complete approach for Java in empirical studies. We found that the assessments 
largely match expert opinion, especially for the maintainability part of the model. 
Yet, the results are not completely conclusive. The model as well as the 
quality assessments were highly understandable for practitioners and considered the
best that can be done with static analysis. The opinions on the validity of the assessments
and on potentially improved productivity were inconclusive, however. 

(5) We developed extensive, 
open-source tool support for building operationalised quality models as well as 
performing the quality assessments including extensive visualisation capabilities.
(6) In addition to the base model, we described the specific quality model for the area
of embedded software. It relies on the same infrastructure as the base model
and shows how the Quamoco approach can be used for a specific context. Furthermore,
it demonstrates the possibility of exploiting the base model to build specific models with
less effort than if starting from scratch.

\subsection{Conclusions}

Overall, we found that it is possible to bridge the gap between abstract quality characteristics 
and concrete measures by building detailed, operationalised quality models. It required great
effort to create the necessary foundations and then build such a detailed model. We are not sure
to what extent such effort can be expended in practical settings. Yet, we found in the 
empirical analyses that there is reasonable agreement between expert opinion and model-based
quality assessments as well as an increased understanding of operationalised quality models by
practitioners.

A measurement program using GQM would probably be able to come up with a similar quality model. It took three years
and about 23 people, however, to build the quality models and surrounding tools and methods. 
most companies are probably not willing to spend this effort but would rather adapt an existing quality model.

There is now a broad, freely available quality model based on an explicit meta-model
capturing the experience of several quality experts from industry and academia. It also
contains complete tool support from model creation to quality assessment. 

For academia,
this allows investigating the model in detail, working on different aspects of the model that
are still under-developed and empirically investigating and comparing the quality model with
other approaches. In that sense, the Quamoco base model can be used as a benchmark.

The base model itself is broad and covers most quality characteristics of ISO/IEC~25010
but with varying degrees of depth. Maintainability, in particular, is well covered and seems to be assessable
by static and manual analyses. Other quality characteristics, such as reliability or security,
are covered by static checks, but we expect that dynamic analyses will be able to improve
the assessments. The modelling of such aspects can be done with the existing mechanisms,
but the execution of quality assessment will be different as we need a running system in a
suitable environment.

In the follow-up of the project, the Quamoco meta-model and tooling were also shown to be ready for industrial use: itestra GmbH deployed them successfully at a financial software provider with the purpose of managing the quality and costs of all larger projects. 
Further commercial applications may follow.
Key features were the flexible modelling of quality characteristics, measures, aggregation and
evaluation rules, as well as the ability to deal with different technologies, such as
Java, .NET, C++, and COBOL in a similar, integrated way. 
We also transferred the approach successfully into practice in a project with several SMEs~\cite{gleirscher14}. In that project,
however, we also found a partial mismatch between the analysis results of the base model and the
opinions of the experts. This suggests that the base model would need to be adapted better to their environment.

We invested a lot of effort in creating a clear and efficient approach for building the quality model.
On the other hand, we learned a lot along the way. We found the issues in Table~\ref{tab:lessons-learned} to be the
most important lessons learned from the Quamoco project. 

\renewcommand{\arraystretch}{2}
\begin{table*}[htbp]
\small
  \caption{Lessons Learned}
  \centering
  \begin{tabular}{R{2.7cm}p{13.2cm}}
    \hline
    \textbf{Additional layer} & Our discussions and experiences led us to the decision to include --
          in contrast to most existing quality models such as ISO/IEC~25010 or \cite{samoladas08} -- an
          additional layer, which we called product factors, in the quality models to make it easier to relate measures
          and quality aspects to it. We can confirm similar proposals made by Kitchenham et al.~\cite{Kitchenham.1997} or
          Dromey~\cite{Dromey.1995}, because this allows clear operationalisation of the model. The \emph{practices}
          of Squale~\cite{MordalManet.2009} go into a similar direction but are less clearly defined.\\
    \hline
    \textbf{Modularisation} & Modularisation proved to be a powerful concept in the models for handling different
	technologies and paradigms and their overlapping quality characteristics. Each module can place special emphasis on
	its specifics and extend and use the elements of (other) existing modules. Modularisation had not
	been used in quality models before.\\ 
    \hline
        \textbf{Operationalisation} & Completely operationalising quality models using measures, instruments and
        evaluations is an elaborate task. We spent many person-months on the model and were still only able to
        cover a few quality aspects and technologies comprehensively. We found in our validations, however, that the
        operationalisation makes the abstract quality 
        attributes from ISO easier to understand and explains software metrics for practitioners. It helps to bridge
        the identified gap. Again, Squale~\cite{MordalManet.2009} is most similar here. They report first results
        of their operationalised quality model being well accepted by developers at Air France, but they did not
        perform a thorough analysis. Hence, it remains unclear if the effort pays off in the end.\\
    \hline
    \textbf{Rankings and threshold calibration} & Determining good thresholds in the assessment is crucial but hard. 
         The benchmarking approach helped us to put this on a more solid foundation, but it is an elaborate process and a 
         large number of available systems are required. Our calibration approach is similar to the benchmarking in 
         \cite{alves11} which we can confirm as a useful means for determining thresholds. Nevertheless, the participants
         of our study still found the threshold determination and calibration hard to understand.\\
    \hline
    \textbf{Static analysis for quality assessment} & Static analysis is extremely handy for software analyses as it needs neither a specific context nor 
    test cases and a running system. Hence, we can easily and frequently run such analyses on many systems. This advantage is
    also a disadvantage because we found that inherently dynamic aspects, such as reliability or security, are difficult to capture
    statically. Nevertheless, we saw mostly reasonable assessments in concordance with expert opinion.\\
   \hline
    \textbf{Nightly quality analysis} & Introducing nightly quality analyses, similar to nightly builds for software, proved to be
	the most important step on the way to an operationalised quality model. We ran the
	current quality model every night on a wide range of systems to detect problems
	in the model and tooling as soon as possible. To the best of our knowledge, such an approach has not been used before. \\ 
    \hline
   \textbf{Difficult validation} & Objective validation of the assessment results is hard because we usually do
	not have objective, independent quality assessments to compare our results to. Our assessment
	results cannot be compared with defect counts, and other general quality metrics are
	usually not available. This leaves us with a low sample size so far. Bansiya and Davis~\cite{Bansiya2002}, for example, 
	used a similar approach to validate their QMOOD model.	\\
    \hline
  \end{tabular}
  \label{tab:lessons-learned}
\normalsize
\end{table*}

\subsection{Future Work}

By working on filling the gap in current quality models, we found several other directions
for future work that should be followed. First, the base model and its validation focus 
on a small number of paradigms and technologies so far. To be truly broad, we need to 
take into account further contents for the base model. Second, we are working on additional 
empirical studies to better understand the remaining weaknesses of our approach to further 
improve it accordingly.

Third, we believe that the tool support is exceptional for a research prototype but not yet
on the level of a commercial or mature open source application. Thus, we will continue to work 
on the tools as an open source project. Fourth, Mordal-Manet et al.~\cite{Mordal.2012} present an in-depth discussion
of possibilities for aggregating quality metrics. This should also be investigated in the context of
Quamoco models and their assessment method. Finally, the majority of the base model measures 
have static analysis instruments to allow them to be collected with low effort. This will, 
however, limit the possibilities of assessing dynamic aspects
of the software. Hence, we are working on extensions to include testing and other
dynamic analyses.

\section*{Acknowledgement}

We are grateful to all members of the Quamoco project team
as well as to all the participants of our interviews and surveys. This work has been
supported by the German Federal Ministry of Education and Research (BMBF) under
grant number 01IS08023.

\bibliographystyle{plain}

\begin{thebibliography}{10}

\bibitem{AlKilidar.2005}
Hiyam Al-Kilidar, Karl Cox, and Barbara Kitchenham.
\newblock The use and usefulness of the {ISO/IEC} 9126 quality standard.
\newblock In {\em Proc.~International Symposium on Empirical Software
  Engineering (ISESE'05)}, pages 7--16. IEEE, 2005.

\bibitem{Alshayeb.2003}
Mohammad Alshayeb and Wei Li.
\newblock An empirical validation of object-oriented metrics in two different
  iterative software processes.
\newblock {\em IEEE Transactions on Software Engineering}, 29(11):1043--1049,
  2003.

\bibitem{alves11}
Tiago~L. Alves, Jos\'{e}~Pedro Correia, and Joost Visser.
\newblock Benchmark-based aggregation of metrics to ratings.
\newblock In {\em Proc.\ IWSM/Mensura 2011}, pages 20--29. IEEE, 2011.

\bibitem{Ayewah2007}
Nathaniel Ayewah, William Pugh, J.~David Morgenthaler, John Penix, and YuQian
  Zhou.
\newblock Evaluating static analysis defect warnings on production software.
\newblock In {\em Proc. 7th ACM SIGPLAN-SIGSOFT Workshop on Program Analysis
  for Software Tools and Engineering (PASTE '07)}, pages 1--8. ACM, 2007.

\bibitem{Bajracharya:2009lr}
Sushil Bajracharya, Joel Ossher, and Cristina Lopes.
\newblock Sourcerer: An internet-scale software repository.
\newblock In {\em Proc.~2009 {ICSE} Workshop on Search-Driven
  Development--Users, Infrastructure, Tools and Evaluation}, pages 1--4. IEEE,
  2009.

\bibitem{bakota11}
Tibor Bakota, P\'{e}ter Heged\=us, P\'eter K\"ortv\'elyesi, Rudolf Ferenc, and
  Tibor Gyim\'othy.
\newblock A probabilistic software quality model.
\newblock In {\em Proc.~27th IEEE International Conference on Software
  Maintenance}, pages 243--252. IEEE, 2011.

\bibitem{Bansiya2002}
Jagdish Bansiya and Carl~G. Davis.
\newblock A hierarchical model for object-oriented design quality assessment.
\newblock {\em IEEE Transactions on Software Engineering}, 28(1):4 --17, 2002.

\bibitem{Barron1996}
F.~Hutton Barron and Bruce~E. Barrett.
\newblock Decision quality using ranked attribute weights.
\newblock {\em Management Science}, 42(11):1515--1523, 1996.

\bibitem{Basili.1996}
Victor~R Basili, Lionel Briand, and Walcelio~L Melo.
\newblock {A validation of object-oriented design metrics as quality
  indicators}.
\newblock {\em IEEE Transactions on Software Engineering}, 22(10):751--761,
  1996.

\bibitem{Benlarbi.2000}
Saida Benlarbi, Khaled {El Emam}, Nishith Goel, and Shesh~N Rai.
\newblock Thresholds for object-oriented measures.
\newblock In {\em Proc.~11th International Symposium on Software Reliability
  Engineering (ISSRE 2000)}, pages 24--38. IEEE, 2000.

\bibitem{Bergel.2009}
Alexandre Bergel, Simon Denier, St\'{e}phane Ducasse, Jannik Laval, Fabrice
  Bellingard, Philippe Vaillergues, and Fran{\c c}oise Balmas.
\newblock {SQUALE - Software QUALity Enhancement}.
\newblock In {\em Proc.~13th European Conference on Software Maintenance and
  Reengineering}, pages 285--288. IEEE, 2009.

\bibitem{Bessey2010}
Al~Bessey, Dawson Engler, Ken Block, Ben Chelf, Andy Chou, Bryan Fulton, Seth
  Hallem, Charles Henri-Gros, Asya Kamsky, and Scott McPeak.
\newblock {A few billion lines of code later}.
\newblock {\em Communications of the ACM}, 53(2):66--75, 2010.

\bibitem{1978_boehmb_software_quality}
Barry~W. Boehm, John~R. Brown, Hans Kaspar, Myron Lipow, Gordon~J. Macleod, and
  Michael~J. Merrit.
\newblock {\em Characteristics of Software Quality}.
\newblock North-Holland, 1978.

\bibitem{Briand.2000}
Lionel Briand, J\"{u}rgen W\"{u}st, John~W. Daly, and D.~Victor Porter.
\newblock Exploring the relationships between design measures and software
  quality in object-oriented systems.
\newblock {\em Journal of Systems and Software}, 51(3):245--273, 2000.

\bibitem{EN50128:2001}
CENELEC.
\newblock {EN 50128: Railway applications - Communications, signaling and
  processing systems - Software for System Safety}, 2001.

\bibitem{10.1109/QSIC.2006.13}
Victor~K.Y. Chan, W.~Eric Wong, and T.F. Xie.
\newblock Application of a statistical methodology to simplify software quality
  metric models constructed using incomplete data samples.
\newblock {\em Proc.\ International Conference on Quality Software (QSIC'06)},
  pages 15--21, 2006.

\bibitem{1989_davis_tam}
Davis~Fred D.
\newblock Perceived usefulness, perceived ease of use, and user acceptance of
  information technology.
\newblock {\em MIS Quarterly}, 13(3):319--340, 1989.

\bibitem{Dandashi.2002}
Fatma Dandashi.
\newblock A method for assessing the reusability of object-oriented code using
  a validated set of automated measurements.
\newblock In {\em Proc.~ACM Symposium on Applied Computing (SAC'02)}, pages
  997--1003. ACM, 2002.

\bibitem{deissenboeck2011quamoco}
Florian Deissenboeck, Lars Heinemann, Markus Herrmannsdoerfer, Klaus Lochmann,
  and Stefan Wagner.
\newblock The {Quamoco} tool chain for quality modeling and assessment.
\newblock In {\em Proc.~33rd International Conference on Software Engineering
  (ICSE'11)}, pages 1007--1009. ACM, 2011.

\bibitem{deissenb:softw08}
Florian Deissenboeck, Elmar Juergens, Benjamin Hummel, Stefan Wagner, Benedikt
  Mas~y Parareda, and Markus Pizka.
\newblock Tool support for continuous quality control.
\newblock {\em IEEE Software}, 25(5):60--67, 2008.

\bibitem{2009_deissenboeckf_purposes_scenarios}
Florian Deissenboeck, Elmar Juergens, Klaus Lochmann, and Stefan Wagner.
\newblock Software quality models: Purposes, usage scenarios and requirements.
\newblock In {\em Proc.~{ICSE} Workshop on Software Quality}, pages 9--14.
  IEEE, 2009.

\bibitem{2006_deissenboeckf_naming}
Florian Deissenboeck and Markus Pizka.
\newblock Concise and consistent naming.
\newblock {\em Software Quality Journal}, 14(3):261--282, 2006.

\bibitem{deissenb:icsm07}
Florian Deissenboeck, Stefan Wagner, Markus Pizka, Stefan Teuchert, and
  Jean-Fran{\c c}ois Girard.
\newblock An activity-based quality model for maintainability.
\newblock In {\em Proc.~IEEE International Conference on Software Maintenance
  (ICSM'07)}, pages 184--193. IEEE, 2007.

\bibitem{Dodgson2000}
J.~Dodgson, M.~Spackman, A.~Pearman, and L.~Phillips.
\newblock Multi-criteria analysis: A manual.
\newblock Technical report, Department of the Environment, Transport and the
  Regions, London, 2000.

\bibitem{Dromey.1995}
R.~Geoff Dromey.
\newblock A model for software product quality.
\newblock {\em IEEE Transactions on Software Engineering}, 21(2):146--162,
  1995.

\bibitem{Edwards1994}
Wards Edwards and F.~Hutton Barron.
\newblock {SMARTS} and {SMARTER}: Improved simple methods for multiattribute
  utility measurement.
\newblock {\em Organizational Behavior and Human Decision Processes},
  60(3):306--325, 1994.

\bibitem{franch03}
Xavier Franch and Juan~Pablo Carvallo.
\newblock Using quality models in software package selection.
\newblock {\em IEEE Software}, 20(1):34--41, 2003.

\bibitem{gleirscher14}
Mario Gleirscher, Dmitriy Golubitskiy, Maximilian Irlbeck, and Stefan Wagner.
\newblock Introduction of static quality analysis in small- and medium-sized
  software enterprises: experiences from technology transfer.
\newblock {\em Software Quality Journal}, 22(3):499--542, 2014.

\bibitem{grady87}
Robert~B. Grady and Deborah~L. Caswell.
\newblock {\em Software Metrics: Establishing a Company-Wide Program}.
\newblock Prentice Hall, 1987.

\bibitem{Gruber2010}
Harald Gruber, Reinhold Pl\"{o}sch, and Matthias Saft.
\newblock On the validity of benchmarking for evaluating code quality.
\newblock In {\em Proc.~IWSM/MetriKon/Mensura 2010}, 2010.

\bibitem{heitlager07}
Ilja Heitlager, Tobias Kuipers, and Joost Visser.
\newblock A practical model for measuring maintainability.
\newblock In {\em Proc.~6th International Conference on Quality of Information
  and Communications Technology (QUATIC'07)}, pages 30--39. IEEE, 2007.

\bibitem{Hovemeyer.2004}
David Hovemeyer and William Pugh.
\newblock {Finding bugs is easy}.
\newblock {\em ACM Sigplan Notices}, 39(12):92, 2004.

\bibitem{Hovemeyer2007}
David Hovemeyer and William Pugh.
\newblock Finding more null pointer bugs, but not too many.
\newblock In {\em Proc. 7th ACM SIGPLAN-SIGSOFT Workshop on Program Analysis
  for Software Tools and Engineering (PASTE '07)}, pages 9--14. ACM, 2007.

\bibitem{IEC61508:2010}
{IEC 61508}.
\newblock Functional safety of electrical/electronical/programmable electronic
  safety-related systems, 2010.

\bibitem{IEC62304:2006}
{IEC 62304}.
\newblock Medical device software -- software life cycle processes, 2006.

\bibitem{2011_iso_standard_25010}
{ISO/IEC 25010:2011}.
\newblock Systems and software engineering -- systems and software quality
  requirements and evaluation ({SQuaRE}) -- system and software quality models,
  2011.

\bibitem{iso25040}
{ISO/IEC 25040:2011}.
\newblock Systems and software engineering -- systems and software quality
  requirements and evaluation ({SQuaRE}) -- evaluation process, 2011.

\bibitem{iso9126-1:2001}
{ISO/IEC TR 9126-1:2001}.
\newblock Software engineering -- product quality -- part 1: Quality model,
  2001.

\bibitem{Jones.2011}
Capers Jones and Oliver Bonsignour.
\newblock {\em The Economics of Software Quality}.
\newblock Addison-Wesley, 2011.

\bibitem{JSF:2005}
JSF.
\newblock {Joint Strike Fighter Air Vehicle C++ Coding Standards for the System
  Development and Demonstration Program}, 2005.

\bibitem{Juergens.2009}
Elmar Juergens, Florian Deissenboeck, Benjamin Hummel, and Stefan Wagner.
\newblock Do code clones matter?
\newblock In {\em Proc.~31st International Conference on Software Engineering
  (ICSE'09)}, pages 485--495. IEEE, 2009.

\bibitem{Kitchenham.1997}
Barbara Kitchenham, Stephen~G. Linkman, Alberto Pasquini, and Vincenzo Nanni.
\newblock The {SQUID} approach to defining a quality model.
\newblock {\em Software Quality Journal}, 6(3):211--233, 1997.

\bibitem{2011_klaes_QM_adaptation}
M.~Kl\"{a}s, C.~Lampasona, and J.~M\"{u}nch.
\newblock Adapting software quality models: Practical challenges, approach, and
  first empirical results.
\newblock In {\em Proc.~37th Euromicro Conference on Software Engineering and
  Advanced Applications}, pages 341--348. IEEE, 2011.

\bibitem{klaes:emse10}
M.~Kl\"{a}s, H.~Nakao, E.~Elberzhager, and J.~Muench.
\newblock Support planning and controlling of early quality assurance by
  combining expert judgment and defect data -- a case study.
\newblock {\em Empirical Software Engineering Journal}, 15(4):423--454, 2010.

\bibitem{2009_klaes_QM_landscape}
Michael Kl\"{a}s, Jens Heidrich, J\"{u}rgen M\"{u}nch, and Adam Trendowicz.
\newblock {CQML} scheme: A classification scheme for comprehensive quality
  model landscapes.
\newblock In {\em Proc.~35th Euromicro Conference on Software Engineering and
  Advanced Applications (SEAA'09)}, pages 243--250. IEEE, 2009.

\bibitem{Klas.2010}
Michael Kl{\"a}s, Constanza Lampasona, Sabine Nunnenmacher, Stefan Wagner,
  Markus Herrmannsd{\"o}rfer, and Klaus Lochmann.
\newblock How to evaluate meta-models for software quality?
\newblock In {\em Proc.\ IWSM/MetriKon/Mensura 2010}, 2010.

\bibitem{Klas.2011}
Michael Kl{\"a}s, Klaus Lochmann, and Lars Heinemann.
\newblock Evaluating a quality model for software product assessments - a case
  study.
\newblock In {\em Tagungsband Software-Qualit\"atsmodellierung und -bewertung
  (SQMB '11)}. TU M\"unchen, 2011.

\bibitem{Laval.2008}
Jannik Laval, Alexandre Bergel, and Stephane Ducasse.
\newblock Assessing the quality of your software with {MoQam}.
\newblock In {\em 2nd Workshop on FAMIX and Moose in Reengineering
  (FAMOOSr'08)}, 2008.

\bibitem{Leveson.2004}
Nancy~G. Leveson.
\newblock The role of software in spacecraft accidents.
\newblock {\em AIAA Journal of Spacecraft and Rockets}, 41:564--575, 2004.

\bibitem{Lochmann2010}
Klaus Lochmann.
\newblock Engineering quality requirements using quality models.
\newblock In {\em Proc.~15th IEEE International Conference on Engineering of
  Complex Computer Systems (ICECCS'10)}, pages 245--246. IEEE, 2010.

\bibitem{Lochmann.2011}
Klaus Lochmann and Lars Heinemann.
\newblock Integrating quality models and static analysis for comprehensive
  quality assessment.
\newblock In {\em Proc.~2nd International Workshop on Emerging Trends in
  Software Metrics (WETSoM'11)}, pages 5--11. ACM, 2011.

\bibitem{luckey10}
Markus Luckey, Andrea Baumann, Daniel M\'endez~Fern\'andez, and Stefan Wagner.
\newblock Reusing security requirements using an extend quality model.
\newblock In {\em Proc.~2010 ICSE Workshop on Software Engineering for Secure
  Systems (SESS'10)}, pages 1--7. ACM, 2010.

\bibitem{Marinescu.2005}
Cristina Marinescu, Radu Marinescu, Retru~Florin Mihancea, Daniel Ratiu, and
  Richard Wettel.
\newblock {iPlasma}: An integrated platform for quality assessment of
  object-oriented design.
\newblock In {\em Proc.~21st IEEE International Conference on Software
  Maintenance}, pages 77--80. IEEE, 2005.

\bibitem{2004_marinescur_oo_analysis}
Radu Marinescu and Daniel Ratiu.
\newblock Quantifying the quality of object-oriented design: The
  factor-strategy model.
\newblock In {\em Proc.~11th Working Conference on Reverse Engineering
  (WCRE'04)}, pages 192--201. IEEE, 2004.

\bibitem{Mayr.2012}
Alois Mayr, Reinhold Pl\"{o}sch, Michael Kl\"{a}s, Constanza Lampasona, and
  Matthias Saft.
\newblock A comprehensive code-based quality model for embedded systems.
\newblock In {\em Proc.~23th IEEE International Symposium on Software
  Reliability Engineering (ISSRE'12)}, pages 281--290. IEEE, 2012.

\bibitem{Mayr.2014}
Alois Mayr, Reinhold Pl\"{o}sch, and Christian K\"{o}rner.
\newblock Objective safety compliance checks for source code.
\newblock In {\em Companion Proceedings of 36th International Conference on
  Software Engineering (ICSE 2014)}, 2014.

\bibitem{McCall.1977}
Jim~A. McCall, Paul~K. Richards, and Gene~F. Walters.
\newblock {\em Factors in Software Quality}.
\newblock National Technical Information Service, 1977.

\bibitem{MISRA-C:2004}
MISRA.
\newblock {MISRA-C 2004 Guidelines for the use of the C language in critical
  systems}, 2004.

\bibitem{MISRA-CPP:2008}
MISRA.
\newblock {MISRA C++} 2008 guidelines for the use of the {C++} language in
  critical systems, 2008.

\bibitem{Mordal.2012}
Karine Mordal, Nicolas Anquetil, Jannik Laval, Alexander Serebrenik, Bogdan
  Vasilescu, and St{\'e}phane Ducasse.
\newblock Software quality metrics aggregation in industry.
\newblock {\em Journal of Software: Evolution and Process}, 25(10):1117--1135,
  2012.

\bibitem{MordalManet.2009}
Karine Mordal-Manet, Fran{\c c}oise Balmas, Simon Denier, St\'{e}phane Ducasse,
  Harald Wertz, Jannik Laval, Fabrice Bellingard, and Philippe Vaillergues.
\newblock The {Squale} model -- a practice-based industrial quality model.
\newblock In {\em Proc.~IEEE International Conference on Software Maintenance
  (ICSM'09)}, pages 531--534. IEEE, 2009.

\bibitem{Plosch.2008}
Reinhold Pl{\"o}sch, Harald Gruber, A.~Hentschel, Christian K{\"o}rner, Gustav
  Pomberger, Stefan Schiffer, Matthias Saft, and S.~Storck.
\newblock The {EMISQ} method and its tool support-expert-based evaluation of
  internal software quality.
\newblock {\em Innovations in Systems and Software Engineering}, 4(1):3--15,
  2008.

\bibitem{ploesch_et_al_2009}
Reinhold Pl\"{o}sch, Harald Gruber, Christian K\"{o}rner, Gustav Pomberger, and
  Stefan Schiffer.
\newblock A proposal for a quality model based on a technical topic
  classification.
\newblock In {\em Tagungsband des 2.~Workshops zur
  Software-Qualit\"atsmodellierung und -bewertung}. TU M\"unchen, 2009.

\bibitem{Ploesch_et_al_2010}
Reinhold Pl\"{o}sch, Alois Mayr, and Christian K\"{o}rner.
\newblock Collecting quality requirements using quality models and goals.
\newblock In {\em Proc.~2010 Seventh International Conference on the Quality of
  Information and Communications Technology (QUATIC'10)}, pages 198--203. IEEE,
  2010.

\bibitem{samoladas08}
Ioannis Samoladas, Georgios Gousios, Diomidis Spinellis, and Ioannis Stamelos.
\newblock The {SQO-OSS} quality model: Measurement based open source software
  evaluation.
\newblock In {\em Open Source Development, Communities and Quality. IFIP 20th
  World Computer Congress}, pages 237--248. Springer, 2008.

\bibitem{Schackmann.2009}
Holger Schackmann, Martin Jansen, and Horst Lichter.
\newblock Tool support for user-defined quality assessment models.
\newblock In {\em Proc.~MetriKon 2009}, 2009.

\bibitem{Shatnawi.2010}
Raed Shatnawi.
\newblock A quantitative investigation of the acceptable risk levels of
  object-oriented metrics in open-source systems.
\newblock {\em IEEE Transactions on Software Engineering}, 36(2):216--225,
  2010.

\bibitem{Trend.2010}
Adam Trendowicz, Michael Kl{\"a}s, Constanza Lampasona, Juergen Muench,
  Christian K{\"o}rner, and Saft Matthias.
\newblock Model-based product quality evaluation with multi-criteria decision
  analysis.
\newblock In {\em Proc.\ IWSM/MetriKon/Mensura 2010}, 2010.

\bibitem{Trendowicz_Kopczynska_2014}
Adam Trendowicz and Sylwia Kopczy\'{n}ska.
\newblock Adapting multi-criteria decision analysis for assessing the quality
  of software products. {C}urrent approaches and future perspectives.
\newblock In {\em Advances in Computers}, pages 153--226. Elsevier, 2014.

\bibitem{Solingen1999}
Rini van Solingen and Egon Berghout.
\newblock {\em Goal/Question/Metric Method}.
\newblock McGraw-Hill Professional, 1999.

\bibitem{vanzeist96}
R.~H.~J. van Zeist and P.~R.~H. Hendriks.
\newblock Specifying software quality with the extended {ISO} model.
\newblock {\em Software Quality Journal}, 5(4):273--284, 1996.

\bibitem{vincke_1992}
P.~Vincke.
\newblock {\em Multicriteria Decision-Aid}.
\newblock Contemporary Evaluation Research. Wiley, 1992.

\bibitem{wagner:ist10}
Stefan Wagner.
\newblock A {Bayesian} network approach to assess and predict software quality
  using activity-based quality models.
\newblock {\em Information and Software Technology}, 52(11):1230--1241, 2010.

\bibitem{wagner:quamoco-data}
Stefan Wagner, Klaus Lochmann, Lars Heinemann, Michael Kl\"{a}s, Constanza
  Lampasona, Adam Trendowicz, Reinhold Pl\"{o}sch, Alois Mayr, Andreas Seidl,
  Andreas Goeb, and Jonathan Streit.
\newblock Questionnaires and raw data for the paper ``{O}perationalised product
  quality models and assessment: The {Q}uamoco approach''.
\newblock http://dx.doi.org/10.5281/zenodo.13290.

\bibitem{wagner:icse12}
Stefan Wagner, Klaus Lochmann, Lars Heinemann, Michael Kl\"{a}s, Adam
  Trendowicz, Reinhold Pl\"{o}sch, Andreas Seidl, Andreas Goeb, and Jonathan
  Streit.
\newblock The {Quamoco} product quality modelling and assessment approach.
\newblock In {\em Proc.~34th International Conference on Software Engineering
  (ICSE'12)}, pages 1133--1142. IEEE, 2012.

\bibitem{Wagner.2012x}
Stefan Wagner, Klaus Lochmann, Sebastian Winter, Florian Deissenboeck, Elmar
  Juergens, Markus Herrmannsdoerfer, Lars Heinemann, Michael Kl{\"a}s, Adam
  Tendowicz, Jens Heidrich, Reinhold Ploesch, Andreas Goeb, Christian Koerner,
  Korbinian Schoder, Jonathan Streit, and Christian Schubert.
\newblock The {Q}uamoco quality meta-model.
\newblock Technical Report TUM-I1281, {T}echnische {U}niversit{\"a}t
  {M}{\"u}nchen, 2012.

\bibitem{2009_wagners_quality_models_practice}
Stefan Wagner, Klaus Lochmann, Sebastian Winter, Andreas Goeb, and Michael
  Klaes.
\newblock Quality models in practice: A preliminary analysis.
\newblock In {\em Proc.~3rd International Symposium on Empirical Software
  Engineering and Measurement (ESEM'09)}, pages 464--467. IEEE, 2009.

\bibitem{Wagner2010}
Stefan Wagner, Klaus Lochmann, Sebastian Winter, Andreas Goeb, Michael
  Kl\"{a}s, and Sabine Nunnenmacher.
\newblock Software quality in practice. survey results.
\newblock Technical Report TUM-I129, Technische Universit\"{a}t M\"{u}nchen,
  2012.

\bibitem{Wicks.2005}
Michael Wicks.
\newblock A software engineering survey.
\newblock Technical Report HW-MACS-TR-0036, {H}eriot-{W}att {U}niversity, 2005.

\end{thebibliography}

\appendix

 \section{Benchmarking Approach}
\label{sec:benchmarking}

In this appendix, we describe the details of how to determine the thresholds \emph{min} 
and \emph{max} for the utility functions in the assessment approach using benchmarking.
We use Equation~\ref{eq:calibration}, where $s_i = S(F_x)$ is the normalised measurement result
for the baseline system $i$ and $Q1$ and $Q3$ represent the 25\% 
and 75\% percentiles; $IQR = Q3 - Q1$ is the inter-quartile range. 

\begin{equation}
\label{eq:calibration}
\scalebox{0.8}{$
\begin{array}{l}\displaystyle
	\mbox{IF } \left| \left\{ s_{i=1...n} : s_i > 0 \right\} \right| < 5 \mbox{ THEN} \\
	\quad min = 0, max = 0.00000001 \\
	\mbox{ELSE}\\
	\quad \begin{array}{ll}max = max&\left(\left\{ 
	s_i : s_i \leq 
	\begin{array}{l}
	 Q3(\{s_i : s_i \neq 0\})
\\+ 1.5 \times IQR(\{s_i\}) 
\end{array}
\right\}\right), \\
min = min&\left(\left\{ 
	s_i : s_i \geq 
	\begin{array}{l}
	 Q1(\{s_i : s_i \neq 0\})
\\- 1.5 \times IQR(\{s_i\}) 
\end{array}
\right\}\right) \\
\end{array}\\
\mbox{END}
\end{array}$
}
\end{equation}

\begin{figure}[htb]
	\begin{tabularx}{\columnwidth}{XXX}
		\includegraphics[width=0.3\columnwidth]{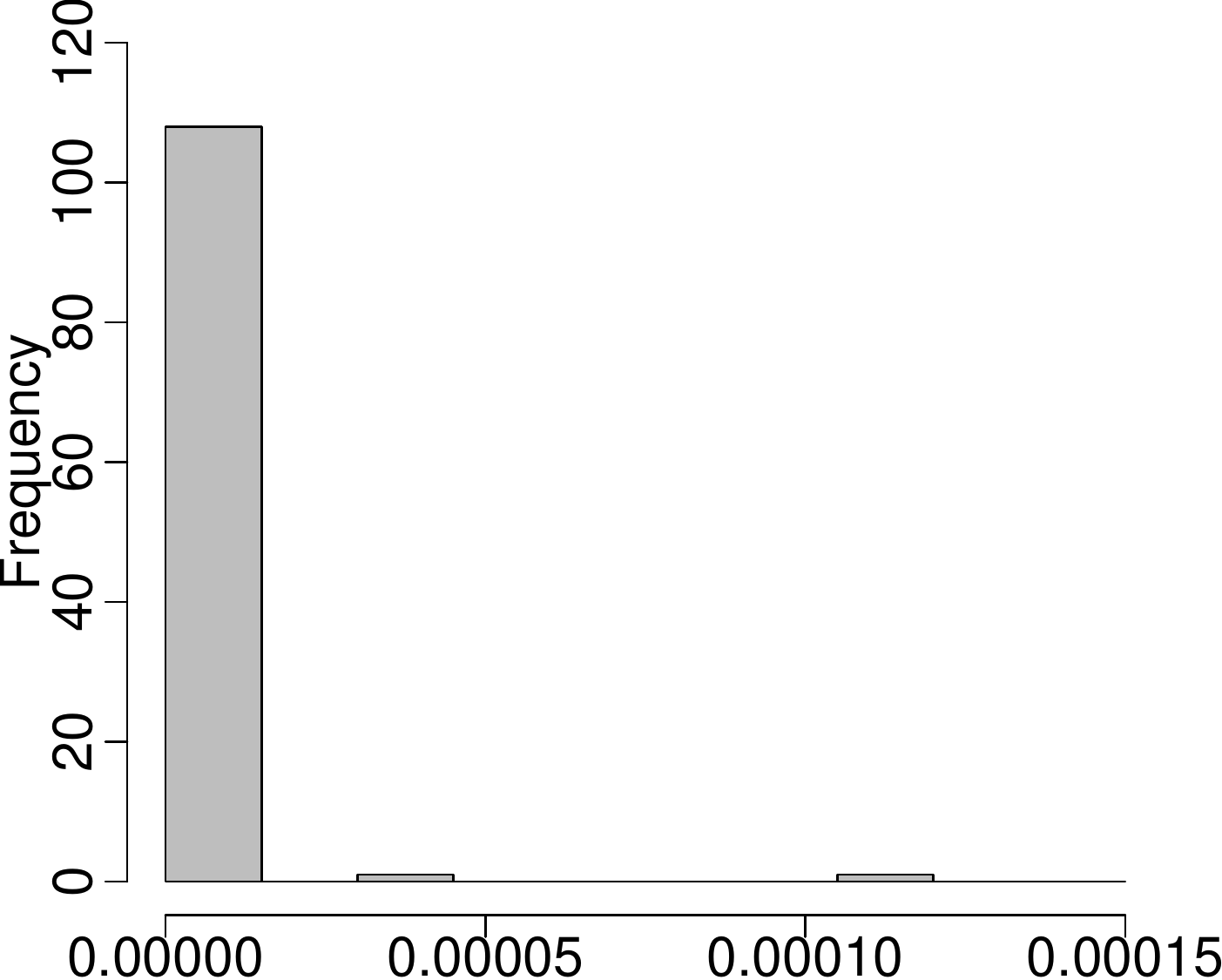}		
		&
		\includegraphics[width=0.3\columnwidth]{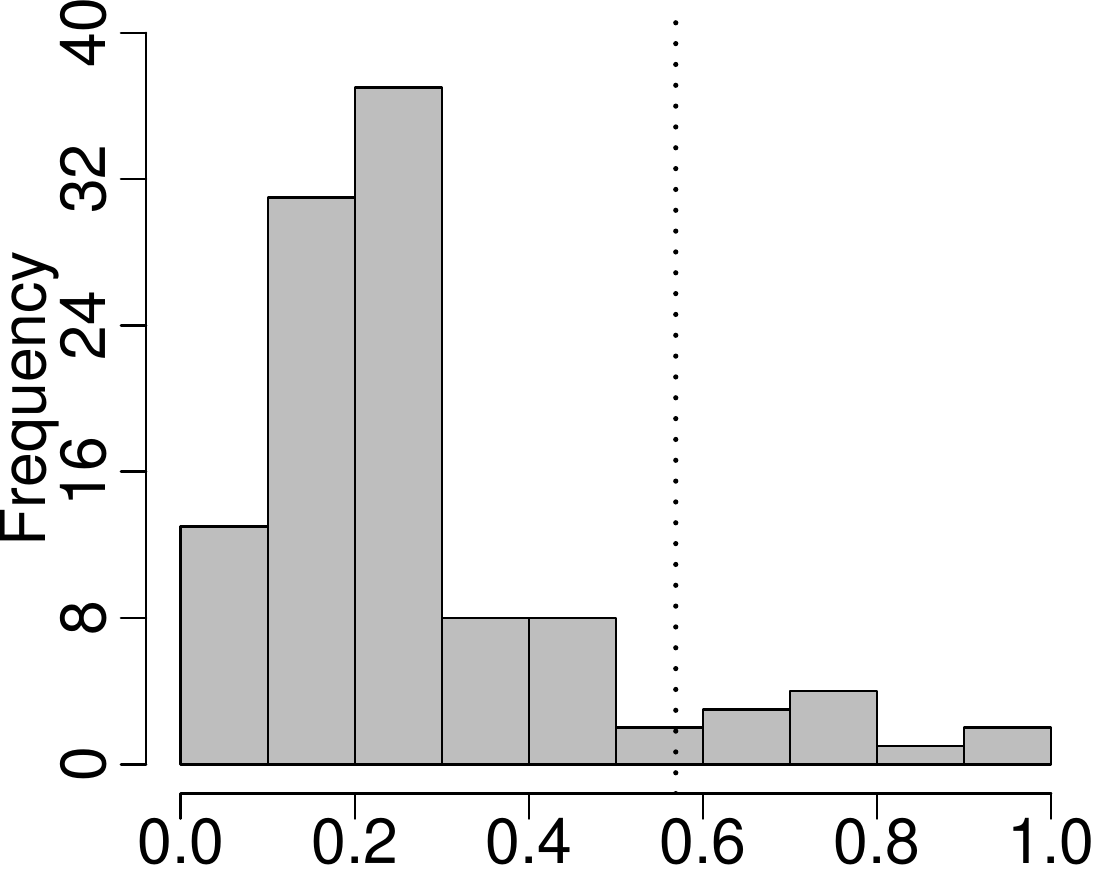}
		&
		\includegraphics[width=0.3\columnwidth]{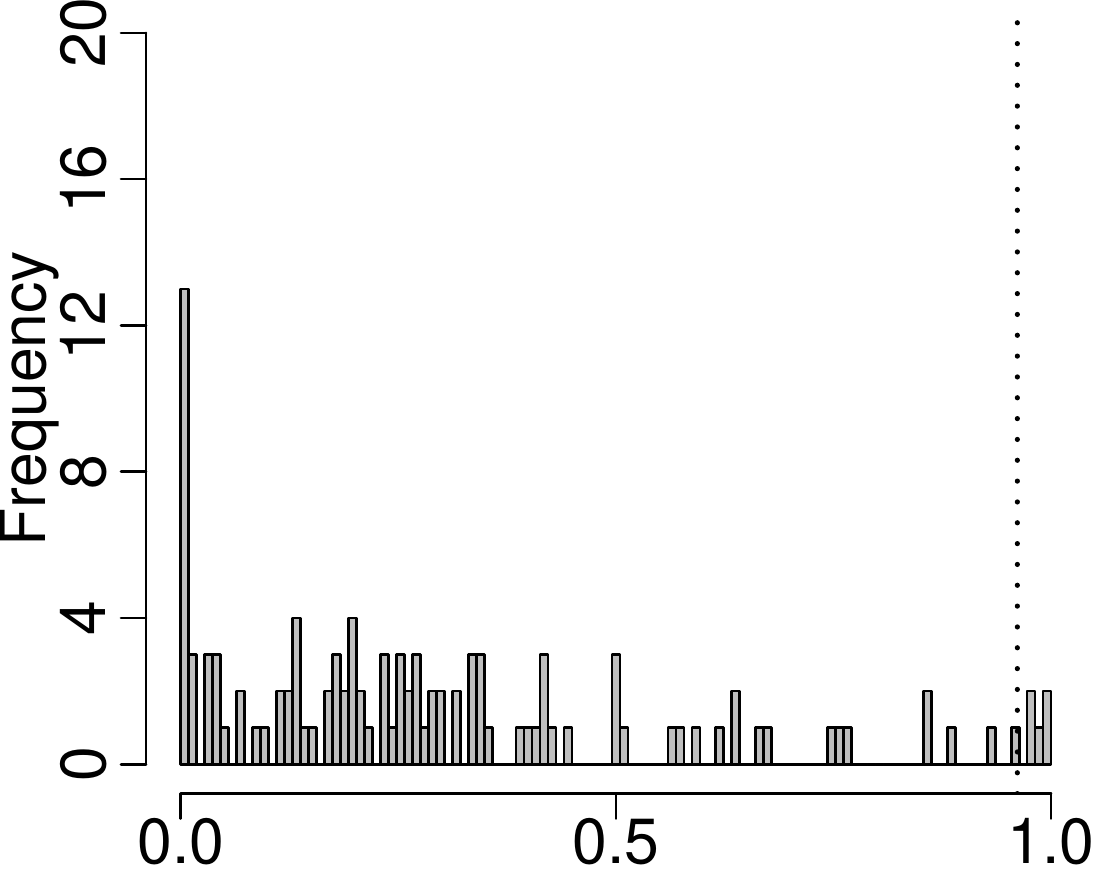}
		\\
		\footnotesize(a) Bitwise Add of Signed Byte
		&
		\footnotesize(b) Clone Coverage 
		&
		\footnotesize(c) Missing JavaDoc Comment \\
	\end{tabularx}
	\caption{Histograms for Three Measures (The values to the right of the dotted lines are considered outliers.)}
	\label{fig:hist:metrics}
\end{figure}

According to this formula, values above $Q3+1.5\times IQR$ (and below $Q1-1.5\times IQR$) 
are considered outliers. This equation takes into consideration typical distributions of measure values:

\begin{enumerate}
\item Figure~\ref{fig:hist:metrics}a shows the histogram of the measure \emph{bitwise add of signed value} 
of 110~systems. This measure is based on a rule-based static code analysis tool and 
normalised by LOC. We can see that all but two systems had no violations of this rule and 
have a normalised value of $0$. Such a distribution frequently appears for rule-based measures. In 
this case, Equation~\ref{eq:calibration} realises a simple jump function at 0. 
\item Figure~\ref{fig:hist:metrics}b shows the histogram of the measure \emph{clone coverage}. 
This metric is not normalised because it returns one value for the entire system, which is
already ``normalised'' by definition. The distribution of values is typical for ``classical'' 
software metrics. In this case, Equation~\ref{eq:calibration} uses the minimum and maximum 
non-outlier values as thresholds, \ie $0.0$ and $0.57$. 
\item Figure~\ref{fig:hist:metrics}c shows the histogram of the measure \emph{missing JavaDoc comment}. 
Although it is a rule-based measure, its normalised values are more equally distributed than in~(1). Thus,
we used the non-outlier values as thresholds. On the left side of the distribution,
there are no outliers. On the right side, all values above the dotted line in the figure
are considered outliers.
\end{enumerate}

The approach ensures that outliers in the measurement data do not lead to extreme \emph{min} and \emph{max} thresholds which would lead to narrowly clustered evaluation results due to a stretched evaluation function and, in consequence, to low entropy and lack of differentiation \cite{Klas.2011}.

\end{document}